\documentclass{aa}
\usepackage{upgreek}
\usepackage{graphicx}
\usepackage[varg]{txfonts}
\usepackage{natbib}
\usepackage{soul}
\newlength{\linwx}
\usepackage{color}

\begin{document}

\title{The growth of planets by pebble accretion in evolving protoplanetary discs}
\author{
Bertram Bitsch \inst{1},
Michiel Lambrechts \inst{1},
\and
Anders Johansen \inst{1}
}
\offprints{B. Bitsch,\\ \email{bert@astro.lu.se}}
\institute{
Lund Observatory, Department of Astronomy and Theoretical Physics, Lund University, 22100 Lund, Sweden
}
\abstract{
The formation of planets depends on the underlying protoplanetary disc structure, which in turn influences both the accretion and migration rates of embedded planets. The disc itself evolves on time scales of several Myr, during which both temperature and density profiles change as matter accretes onto the central star. Here we used a detailed model of an evolving disc to determine the growth of planets by pebble accretion and their migration through the disc. Cores that reach their pebble isolation mass accrete gas to finally form giant planets with extensive gas envelopes, while planets that do not reach pebble isolation mass are stranded as ice giants and ice planets containing only minor amounts of gas in their envelopes. Unlike earlier population synthesis models, our model works without any artificial reductions in migration speed and for protoplanetary discs with gas and dust column densities similar to those inferred from observations. We find that in our nominal disc model, the emergence of planetary embryos preferably tends to occur after approximately 2 Myr in order to not exclusively form gas giants, but also ice giants and smaller planets. The high pebble accretion rates ensure that critical core masses for gas accretion can be reached at all orbital distances. Gas giant planets nevertheless experience significant reduction in semi-major axes by migration.  Considering instead planetesimal accretion for planetary growth, we show that formation time scales are too long to compete with the migration time scales and the dissipation time of the protoplanetary disc. All in all, we find that pebble accretion overcomes many of the challenges in the formation of ice and gas giants in evolving protoplanetary discs.
}
\authorrunning{Bitsch et al.}\maketitle

\section{Introduction}
\label{sec:introduction}

The formation of planets takes place in protoplanetary discs that surround newly born stars. This process can happen on different time scales. Gas giants have to form within the lifetime of the gaseous protoplanetary disc, since they must accrete a gaseous envelope after the formation of the planetary core. Because the typical disc lifetimes are constrained to a few Myr \citep{1998ApJ...495..385H, 2001ApJ...553L.153H, 2009AIPC.1158....3M}, giant planet formation has to happen within the same time span. Terrestrial planet formation, on the other hand, finishes on much longer time scales. In our own solar system the last giant impact is constrained to have occurred $\sim~100$ Myr after the solar system formed \citep{2009GeCoA..73.5150K, 2014Natur.508...84J}.

The evolution of the protoplanetary disc is crucial for the formation of giant planets, because the whole growth process from dust to planetary cores happens in the gas phase of the disc. The structure of the disc can be approximated by a minimum mass solar nebular (MMSN), in which the gas surface density $\Sigma_{\rm g}$, temperature $T$, and aspect ratio $H/r$ are set to be simple power laws \citep{1977Ap&SS..51..153W, 1981PThPS..70...35H}. However, the structure of the inner regions of protoplanetary discs ($r<10$~AU) is in reality much more complicated than a simple power law, and the disc instead features bumps and dips caused by transitions in opacity \citep{2014A&A...564A.135B, 2014arXiv1411.3255B, 2015arXiv150303352B}. In \citet{2014arXiv1411.3255B} we ran an extensive suite of protoplanetary disc models and provided analytical fitting models of the disc structure that greatly improve on the simplified MMSN model.

We now review some critical processes for how planets are formed and how they are influenced by the structure of the protoplanetary disc.

{\it Planetesimal formation} can be triggered through the streaming instability \citep{2005ApJ...620..459Y, 2007ApJ...662..627J} or in vortices \citep{2015arXiv150105364R}. For this process to happen, dust particles in the protoplanetary disc first have to grow to pebbles by coagulation and condensation \citep{2010A&A...513A..57Z, 2012A&A...539A.148B, 2013A&A...552A.137R}. These pebbles then move radially towards the start owing to gas drag \citep{1977MNRAS.180...57W, 2008A&A...487L...1B}. During their motion, a swarm of pebbles can undergo a gravitational collapse and form a planetesimal, which are the first building blocks of planets. The formation process of planetesimals via the streaming instability is a strong function of the pressure gradient in the disc, making discs with bumps in their pressure profile very appealing compared to the simple MMSN disc \citep{2014arXiv1411.3255B}.

{\it The growth of the core} of a giant planet can happen via the accretion of planetesimals onto a planetary embryo, which is basically a large planetesimal. However, this process can easily take longer than the lifetime of the protoplanetary disc itself \citep{1996Icar..124...62P, 2004AJ....128.1348R, 2010AJ....139.1297L}, if the amount of solids is not increased by a factor of $6-8$ compared to solar value. These growth time scales can nevertheless be significantly reduced when the accretion of small pebbles onto planetesimals is taken into account \citep{2010MNRAS.404..475J, 2010A&A...520A..43O, 2012A&A...544A..32L, 2012A&A...546A..18M}. In this process the formation of a core of $\sim 10 {\rm M}_{\rm Earth}$ at $\sim 5$~AU can occur within $1$~Myr. Pebbles can either form by coagulation of dust in the disc or be the result of fragmentation of the planetesimal population \citep{2014Icar..233...83C}. Recently, \citet{Johansen2015} have also shown that even chondrules can be accreted effectively on planetesimal seeds. The process of pebble accretion depends significantly on the disc structure because the accretion rate depends on the disc's aspect ratio and pressure gradients \citep{2014arXiv1411.3255B}.

{\it Gas accretion} starts after the core has reached its isolation mass for either planetesimals or pebbles, because during the accretion of solids the atmosphere is heated by the impacts, preventing a contraction and efficient gas accretion. This isolation mass depends on the semi-major axis and column density of planetesimals \citep{1980PThPh..64..544M, 2002ApJ...581..666K, 2014prpl.conf..595R} and on the disc's aspect ratio $H/r$ for pebble accretion \citep{2014arXiv1408.6087L}. After the bombardment of planetesimals or pebbles has stopped, the gaseous envelope can contract, and runaway gas accretion can start \citep{1996Icar..124...62P}. However, the isolation mass of planetesimals is very hard to determine and much clearer to determine for pebble accretion. For planetesimal accretion, the isolation mass has to be estimated by N-body simulations \citep{2002ApJ...581..666K}, while for pebble accretion the isolation mass is determined directly by the modifications of the pressure gradient in the disc, which is caused by the planet itself, and that can halt pebble accretion \citep{2006A&A...453.1129P, 2014arXiv1408.6087L}. The gas accretion is limited not only by the properties of the planet (e.g. mass of the core), but also by the disc itself, because the planet cannot accrete more gas than what is provided through the accretion rate of the gas onto the central star. In fact, the accretion rate onto the planet is at its maximum roughly $\sim 80\%$ of the stellar accretion rate \citep{2006ApJ...641..526L}.

{\it Planet migration} describes the gravitational interactions of the planet with the surrounding gas disc \citep{1997Icar..126..261W}. In the locally isothermal limit, the time scales for inward migration of embedded planets is shorter than the disc's lifetime ($\tau_{\rm mig} \approx 8 \times 10^5$ yr for an Earth size planet at $5$~AU), which poses a problem for the formation of planets \citep{2002ApJ...565.1257T}. Considering the thermodynamics inside the disc, recent studies have shown that planets can migrate outwards \citep{2006A&A...459L..17P, 2008A&A...487L...9K, 2008ApJ...672.1054B, 2009A&A...506..971K}. This outward migration depends on the gradient of entropy in the disc and is most likely to happen in regions of the disc where $H/r$ drops with increasing orbital distance \citep{2013A&A...549A.124B}. Low-mass planets are in this so-called type-I-migration phase, where the perturbation of the planet onto the disc is small. When the planets become more massive, for example, when it is caused by gas accretion, they start to open a gap inside the disc and migrate with the viscous accretion speed of the disc, which is much slower than the type-I migration and is called type-II migration \citep{1986ApJ...309..846L}. Migration thus affects planets of all masses, where its effects become significant when the planet is larger than one Earth mass.

All these aspects and their interplay have to be considered when trying to explain the observed distribution of planets and exoplanets. First attempts to explain the distribution of exoplanets have been done in so called {\it \emph{population synthesis}} studies that started about a decade ago \citep{2004ApJ...604..388I, 2004A&A...417L..25A}. These studies generally assume that the core grows via the accretion of planetesimals, after which gas accretion can set in. During this growth phase, the planets migrate through the disc. These models are able to explain the distribution of the observed exoplanets only by making some critical assumptions, some of which are questionable. These questionable assumptions regarding the migration speed of planets, the amounts of solids in the disc and the lifetime and evolution of protoplanetary disc itself are discussed in more detail in  section~\ref{subsec:popsynth}, where we show that no supposedly helpful assumptions have to be made in our model.

The aim of this paper is to study the formation and evolution of planets in evolving accretion discs around young stars, where planets first grow via pebble accretion and can then contract a gaseous envelope. We focus here on the formation of different planetary types that can emerge in protoplanetary discs, on the parameters in initial semi-major axis, and on the initial time needed to form planets of a certain planetary type. 

\citet{2014arXiv1408.6087L} and \citet{2014arXiv1408.6094L} propose that the dichotomy between ice and gas giants is a natural consequence of fast growth by pebbles and the existence of a pebble isolation mass. Additionally, \citet{2014arXiv1408.6094L} show that ice giants can overcome the type-I migration barrier by growing faster than they migrate. In this study we wish to investigate this concept in a more realistic disc than an MMSN, compared to their study improved planet migration and gas accretion rates.

We use the disc evolution model of \citet{2014arXiv1411.3255B} for solar-type stars. This disc model is a semi-analytical formula fitted to 2D radiation hydrodynamic simulations that feature stellar and viscous heating, as well as radiative cooling. It reproduces the dips and bumps in the disc profile caused by opacity transitions and captures the disc evolution on a time scale of several Myr, which are linked to observations of accretion discs \citep{1998ApJ...495..385H}. In this disc we implant planetesimals that accrete pebbles, following the radial-drift-dominated approach of \citet{2012A&A...539A.148B, 2014arXiv1408.6094L} for the formation of pebbles. These planets grow rapidly and can reach their pebble isolation mass \citep{2014arXiv1408.6087L} in several 100 kyr, which is when their gas accretion starts. The gas accretion is modelled by using accretion rates of \citet{2010MNRAS.405.1227M} and envelope contraction rates following \citet{2014ApJ...786...21P}. During their growth, the planets migrate through the disc. We use the analytical torque formula for type-I migration of \citet{2011MNRAS.410..293P} to mimic their motion in the disc. When planets become massive and start to open up a gap in the disc, they migrate with the viscous type-II migration \citep{1986ApJ...309..846L}, which is slower than type-I migration. 

Our work is structured as follows. In section~\ref{sec:methods} we explain the different methods used for pebble and gas accretion, for the disc evolution, and for planetary migration. We then present results of simulations where the planets grow via pebble and gas accretion while they migrate through the evolving disc (section~\ref{sec:formation}). The results obtained with pebble accretion are compared with simulations where the cores grow via planetesimal accretion in section~\ref{sec:planetsimals}. In section~\ref{sec:solarsystem} we discuss the formation of the giant planets in our own solar system via pebble accretion. The many applications of our planetary growth model are discussed in section~\ref{sec:discussion}. We finally summarize in section~\ref{sec:summary}.

\section{Methods}
\label{sec:methods}

The methods used in this work are explained in much more detail in the literature cited in the following paragraphs. This section only intends to summarize the methods in a condensed way, so that it is easy to understand the principles on which our work is based. During the disc evolution in time (section~\ref{subsec:discevolve}), planets grow first via pebble accretion (section~\ref{subsec:pebbleaccrete}) very quickly. After they have reached their pebble isolation mass, gas can accrete onto the planet (section~\ref{subsec:gasaccrete}). During the whole growth process planets migrate through the disc, which changes their semi major axes (section~\ref{subsec:migration}).

\subsection{Evolution and structure of the disc in time}
\label{subsec:discevolve}

The lifetime of protoplanetary discs spans 1-10 Myr \citep{1998ApJ...495..385H, 2001ApJ...553L.153H, 2009AIPC.1158....3M}. During the lifetime of the disc, the accretion rate $\dot{M}$ changes in time following constraints from observations of slightly sub-solar mass stars in the Taurus cluster \citep{1998ApJ...495..385H},
\begin{equation}
\label{eq:harttimenew}
 \log \left( \frac{\dot{M}}{M_\odot /\text{yr}} \right) = -8.00 - 1.40  \log \left( \frac{t+10^5\text{yr}}{10^6 \text{yr}} \right) \ .
\end{equation}
The accretion rate $\dot{M}$ can then be related to the viscosity $\nu$ and the gas surface density $\Sigma_{\rm g}$ via 
\begin{equation}
 \dot{M} = 3 \pi \nu \Sigma_{\rm g} \ ,
\end{equation}
where we assume a constant accretion rate for each orbital distance. For the viscosity, we take the $\alpha$ approach \citep{1973A&A....24..337S} with $\alpha=0.0054$ constant throughout the whole disc, where $\nu = \alpha H^2 \Omega_{\rm K}$.

\citet{2014arXiv1411.3255B} calculated the structure of accretion discs around solar type stars with 2D simulations that includ viscous and stellar heating, as well as radiative cooling for several different $\dot{M}$ rates, which correspond to different evolution times of the disc (Eq.~\ref{eq:harttimenew}). We note here that in the \citet{2014arXiv1411.3255B}  disc model, the $\alpha$ value only represents the heating of the disc and is not representative of the viscous evolution of the disc. They then provided a semi-analytical fit to the disc structure evolution in time (see Appendix A in \citet{2014arXiv1411.3255B}), which we use for the evolution model of our disc. This model covers a radial extent from $1$ to $50$~AU. Inside of one, we extrapolate the fit of \citet{2014arXiv1411.3255B} with the given power laws. 

This extension of the disc structure fit is correct as long as the temperature in the disc is so low that silicates do not melt or evaporate. The melting or evaporation causes an additional transition in the opacity profile, which changes the cooling properties of the disc and therefore the structure of the disc. In the very early stages of the disc, the silicate evaporation line is at $0.7$~AU, but it moves inwards in time as the disc loses mass, so that silicates only evaporate in the very inner regions of the disc ($r<0.1$~AU). In addition, we focus on planets that form in the outer disc ($r_{\rm P} > 3$~AU), which only reach the inner regions of the disc via migration when they have stopped accreting solids (Eq.~\ref{eq:Misolation}) in the first place.

The disc structure of \citet{2014arXiv1411.3255B} features bumps and wiggles in the important disc quantities ($\Sigma_{\rm g}, T$, and $H$), which are caused by transitions in the opacity $\kappa$ (e.g. at the ice line) that influences the cooling rates of the disc as $D\propto 1/\kappa$ \citep{2013A&A...549A.124B,2014A&A...564A.135B}. A change in the cooling rate of the disc directly changes the discs temperature $T$ and thus the scale height of the disc [$T \propto (H/r)^2$], which in turn changes the local viscosity of the disc. This change in the local viscosity has to be compensated for by a change in the surface density $\Sigma_{\rm g}$ to have the same $\dot{M}$ at all orbital distances, thus creating a change in the local radial gradient in surface density and pressure $P$ \citep{2014A&A...564A.135B}. Therefore a steeper gradient in temperature will result in a shallower gradient in surface density at the same orbital location.

This has important consequences for the accretion of pebbles (section~\ref{subsec:pebbleaccrete}), which depends on the pressure gradient parameter $\eta$ (eq.~\ref{eq:eta}) and for the migration of planets, which depends on the gradients of temperature, surface density, and entropy (section~\ref{subsec:migration}).

The disc structure in itself depends on the dust grains inside the discs, because those grains are responsible for the absorption and re-emission of photons that distribute the heat inside the disc. The main contribution to the dust opacities originates in micrometre-sized dust grains. Larger dust grains only contribute minimally to the opacity, so that we do not take their contribution into account. We assume here a metallicity of micrometre-sized dust grains of $0.5\%$ or $0.1\%$ of the gas density at all time. We also make the assumption that this small dust is coupled perfectly to the gas and does not evolve its size distribution in time. Here we use the opacity table of \citet{1994ApJ...427..987B}.

The decay of the disc accretion rate from $\dot{M}=1 \times 10^{-7} M_\odot$/yr down to $\dot{M}=1 \times 10^{-9} M_\odot$/yr takes $5$ Myr in \citet{1998ApJ...495..385H}. Using the time evolution of $\dot{M}$ via eq.~\ref{eq:harttimenew}, the disc spends 2 Myr decaying from $\dot{M}=2 \times 10^{-9} M_\odot$/yr to $\dot{M}=1 \times 10^{-9} M_\odot$/yr. However, for these low accretion rates, photoevaporation becomes very efficient, and the disc can dissipate in much shorter time scales \citep{2013arXiv1311.1819A}. For this reason, our nominal disc lifetime is set to $3$ Myr, which is when we assume that photoevaporation clears the disc immediately, but we follow the decay rate of $\dot{M}$ given by eq.~\ref{eq:harttimenew} down to $\dot{M}=2 \times 10^{-9} M_\odot$/yr.

The disc structure significantly changes as the disc evolves in time and as $\dot{M}$ decreases. As the disc reduces in $\dot{M}$ and $\Sigma_{\rm g}$, the disc becomes colder, because viscous heating decreases, which implies that the opacity transition at the ice line moves inwards. This means that the bumps and wiggles in the disc structure ($T$, $\Sigma_{\rm g}$ and $H$) move inwards as well. The star also evolves and changes its luminosity, changing the amount of stellar heating received by the disc and thus changing the temperature, which is all taken into account in the \citet{2014arXiv1411.3255B} model. These changes to the disc structure influence the formation and migration (see Fig.~\ref{fig:Migcont}) of growing protoplanets significantly.

\subsection{Growth via pebbles}
\label{subsec:pebbleaccrete}

The growth of planetary embryos via pebble accretion is outlined in \citet{2012A&A...544A..32L} and \citet{2014arXiv1408.6094L}. The pebbles form from grains initially embedded in the protoplanetary disc ($\sim$ $\upmu$m size) by collisions \citep{2012A&A...539A.148B} or through sublimation and condensation cycles around ice lines \citep{2013A&A...552A.137R}. Swarms of these pebbles drift inwards towards the star, but can collapse under their own gravity and form planetesimals of $100$ to $1000$ km in size in a process called streaming instability \citep{2005ApJ...620..459Y,2007ApJ...662..627J}. Further discussion on this process can be found in the review of \citet{Johansen2014}, along with a list of other models of planetesimal formation by particle concentration and gravitational collapse.

We now consider cores that predominantly grow by accretion of particles with approximately mm-cm sizes. This particle size can be expressed through the gas drag time scale $t_{\rm f}$ and the Keplerian frequency $\Omega_{\rm K}$ in terms of the Stokes number
\begin{equation}
 \label{eq:Stokesnumber}
 \tau_{\rm f} = \Omega_{\rm K} t_{\rm f} = \frac{\rho_\bullet R}{\rho_{\rm g} H_{\rm g}} \ ,
\end{equation}
where $\rho_\bullet$ is the solid density, $R$ the particle radius, $\rho_{\rm g}$ the gas density, $\Omega_{\rm K}$ the Keplerian frequency, and $H_{\rm g}$ the local gas scale height. Small particles ($\tau_{\rm f} \ll 1$) are strongly coupled and move with the gas, while larger particles ($\tau_{\rm f} \gg 1$) are only weakly affected by gas drag. 

The scale height of pebbles $H_{\rm peb}$ is related to the scale height of the gas $H_{\rm g}$ through the viscosity and the Stokes number \citep{2007Icar..192..588Y} by
\begin{equation}
\label{eq:Hpebble}
 H_{\rm peb} = H_{\rm g} \sqrt{\alpha / \tau_{\rm f}} \ ,
\end{equation}
where $\alpha$ is the viscosity parameter. In our simulations we place seed masses that have reached the pebble transition mass 
\begin{equation}
\label{eq:Mtrans}
 M_{\rm t} = \sqrt{\frac{1}{3}} \frac{(\eta v_{\rm K})^3}{G \Omega_{\rm K}} \ ,
\end{equation}
where $G$ is the gravitational constant, $v_{\rm K}= \Omega_{\rm K} r$, and
\begin{equation}
\label{eq:eta}
 \eta = - \frac{1}{2} \left( \frac{H}{r} \right)^2 \frac{\partial \ln P}{\partial \ln r} \ .
\end{equation}
Here, $\partial \ln P / \partial \ln r$ is the radial pressure gradient in the disc. These masses are typically in the range of $5\times 10^{-4} {\rm M}_{\rm E}$ to $10^{-2} {\rm M}_{\rm E}$ (see Fig.~\ref{fig:Envhockey}). The pebble transition mass defines the planetary mass at which pebble accretion occurs within the Hill radius, while for lower masses pebbles are accreted within the Bondi radius \citep{2012A&A...544A..32L}.

These masses are a bit higher than planetesimals formed by the streaming instability, which have roughly $10^{-4}$ ${\rm M}_{\rm E}$ \citep{2012A&A...537A.125J}. Even if a planetesimal of  $10^{-4}$ ${\rm M}_{\rm E}$ forms at $t=0$, the planetesimal has several Myr to grow to the pebble transition mass $M_{\rm t}$. This growth phase can occur through the accretion of planetesimals or pebbles in the inefficient Bondi accretion regime \citep{2012A&A...544A..32L, Johansen2015}.

Planets whose Hill radius is roughly larger than the scale height of the pebbles (eq.~\ref{eq:Hpebble}) accrete in a 2D fashion, and the accretion is given by
\begin{equation}
\label{eq:Mdotpebble}
 \dot{M}_{\rm c, 2D} = 2 \left(\frac{\tau_{\rm f}}{0.1}\right)^{2/3} r_{\rm H} v_{\rm H} \Sigma_{\rm peb} \ ,
\end{equation}
where $r_{\rm H} = r [M_{\rm c} / (3M_\star)]^{1/3}$ is the Hill radius, $v_{\rm H}=\Omega_{\rm K} r_{\rm H}$  the Hill speed, and $\Sigma_{\rm peb}$  the pebble surface density. If the Stokes number of the particles $\tau_{\rm f}$ is larger than $0.1,$ the accretion rate is limited to
\begin{equation}
 \dot{M}_{\rm c, 2D} = 2 r_{\rm H} v_{\rm H} \Sigma_{\rm peb} \ ,
\end{equation}
because the planetary seed cannot accrete particles from outside its Hill radius \citep{2012A&A...544A..32L}. However, when the planets are small and their Hill radius is smaller than the scale height of the pebbles, the pebble accretion rate is reduced and planets accrete in a 3D way, which is related to the 2D accretion rate \citep{Morby2015} by
\begin{equation}
 \dot{M}_{\rm c, 3D} = \dot{M}_{\rm c, 2D} \left( \frac{\pi (\tau_{\rm f}/0.1)^{1/3} r_{\rm H}}{2 \sqrt{2 \pi} H_{\rm peb}} \right) \ . 
\end{equation}
The transition from 3D to 2D pebble accretion is then reached \citep{Morby2015} when
\begin{equation}
 \label{eq:2D3D}
 \frac{\pi (\tau_{\rm f}/0.1)^{1/3} r_{\rm H}}{2 \sqrt{2 \pi}} > H_{\rm peb} \ .
\end{equation}
This transition depends on particle size ($\tau_{\rm f}$) and on the scale height of the disc. This means that in the outer parts of the disc, where $H_{\rm peb}$ is larger and $\tau_{\rm f}$ is smaller, a higher planetary mass is needed to reach the faster 2D pebble accretion branch. We use the Stokes number of the dominant particle size 
\begin{equation}
 \tau_{\rm f} = \frac{\sqrt{3}}{8} \frac{\epsilon_{\rm P}}{\eta} \frac{\Sigma_{\rm peb}}{\Sigma_{\rm g}} \ .
\end{equation}
This size is obtained from an equilibrium between growth and drift to fit constraints from advanced coagulation models and observations of pebbles in protoplanetary discs \citep{2012A&A...539A.148B}. The parameter $\epsilon_{\rm P}$ is  $0.5$ and $\epsilon_{\rm D}$ is $0.05$ \citep{2014arXiv1408.6094L}. In our disc model, this results in Stokes numbers between $0.05$ and $0.5$. The pebble surface density depends on the gas surface density $\Sigma_{\rm g}$ and the semi major axis $r_{\rm p}$ of the planet through
\begin{equation}
 \label{eq:SigmaPeb}
 \Sigma_{\rm peb} = \sqrt{\frac{2 \dot{M}_{\rm peb} \Sigma_{\rm g} }{\sqrt{3} \pi \epsilon_{\rm P} r_{\rm P} v_{\rm K}}} \ ,
\end{equation}
where the pebble flux is
\begin{equation}
 \dot{M}_{\rm peb} = 2 \pi r_{\rm g} \frac{dr_{\rm g}}{dt} (Z \Sigma_{\rm g}) \ .
\end{equation}
Here, $Z$ denotes the fraction of solids (metallicity) in the disc that can be transformed into pebbles at the pebble production line $r_{\rm g}$ at time $t$ 
\begin{equation}
 r_{\rm g} = \left(\frac{3}{16}\right)^{1/3} (GM_\star)^{1/3} (\epsilon_{\rm D} Z)^{2/3} t^{2/3} \ ,
\end{equation}
and
\begin{equation}
 \frac{dr_{\rm g}}{dt} = \frac{2}{3} \left(\frac{3}{16}\right)^{1/3} (GM_\star)^{1/3} (\epsilon_{\rm D} Z)^{2/3} t^{-1/3} \ ,
\end{equation}
where $M_\star$ is the stellar mass, which we set to $1 M_\odot$. After $3$ Myr of disc evolution, the pebble production line is at $250$~AU, indicating that the disc has to be at least $250$~AU wide to sustain a pebble flux for $3$ Myr. After $5$ Myr of disc evolution, the pebble production line is located at $360$~AU. Observations by \citet{2010ApJ...723.1241A} find typical protoplanetary disc radii to be $150-200$~AU for discs that are a few Myr old, which is only slightly smaller than our estimated pebble production line at $3$ Myr.

When pebbles form in the outer disc drift across the water ice line, they will melt, because they mainly consist of ice, and release the trapped silicate particles. Then the particle size shrinks significantly, which will slow down the pebble accretion rate onto the planet (eq.~\ref{eq:Mdotpebble}). However, in our model, the water ice line is located outside $3$~AU only in the very early stages of the disc evolution, and it moves inwards very quickly with time \citep{2014arXiv1411.3255B}. Since planets grow rapidly by accretion of pebbles locally and experience most migration after reaching pebble isolation mass, the planets only accrete icy pebbles and reach their pebble isolation mass before they migrate across the water ice line.

The initial planetary seeds efficiently accrete pebbles (eq.~\ref{eq:Mdotpebble}) and grow very fast. During this growth process, the planetary seeds can also attract a gaseous envelope (section~\ref{subsec:gasaccrete}). The structure of the gaseous envelope is supported by the accretion luminosity deposited by the accreted pebbles into the atmosphere of the planet. The envelope will collapse when the mass of the core is similar to the mass of the envelope itself. However, when the envelope collapses, this critical core mass is a function of the accretion rate onto the planet itself, where the critical core mass becomes higher with increasing accretion rates. Without any interruption in the pebble accretion rate, the critical core mass can be up to $\sim 100 {\rm M}_{\rm E}$, which is up to an order of magnitude higher than the amount of solids in the giant planet's cores of the solar system \citep{2014arXiv1408.6087L}.

However, when the planet reaches a certain mass, it changes the gas pressure gradient in the disc locally, which modifies the rotation velocity of the gas that then halts the drift of pebbles that could be accreted onto the core so the accretion of pebbles stops \citep{2006A&A...453.1129P, 2014arXiv1408.6087L}. This is the so-called pebble isolation mass \citep{2014arXiv1408.6087L}
\begin{equation}
\label{eq:Misolation}
 M_{\rm iso} \approx 20  \left( \frac{H/r}{0.05}\right)^3 M_{\rm Earth} \ .
\end{equation}
When pebble isolation mass is reached, the planet can contract its envelope and start gas accretion. The pebble isolation mass is therefore a natural division between gas and ice giants, where ice giants did not reach pebble isolation mass early in the disc lifetime. This can then explain the difference between gas and ice giants in our own solar system \citep{2014arXiv1408.6087L}.

\subsection{Gas accretion}
\label{subsec:gasaccrete}

Planets that reach pebble isolation mass (eq.~\ref{eq:Misolation}) can start to accrete gas, because the pebble flux onto the planet that heats the envelope and hinders its contraction has stopped.
However, during the formation of the planetary cores via pebble accretion, small amounts of highly polluted gas can be bound to the planet inside the planets Hill sphere. We therefore assume that $10\%$ of the planets mass is in gas, prior to the point when the planet reaches pebble isolation mass. This means that $90\%$ of the nominal pebble accretion rate is counted as solids, and $10\%$ is counted as gas. During this stage the planet grows at the same speed as if $100\%$ of the accretion were in pebbles. This approach also means that the same pebble isolation mass is reached, because the total mass of the planet is the same. 

\citet{2014ApJ...786...21P} estimate the gas contraction time of a gaseous envelope around a planet. After the planet has reached pebble isolation mass, the envelope of the planet contracts on a long time-scale while it accretes some gas. This contraction phase takes place as long as $M_{\rm env} < M_{\rm core}$ and the corresponding gas accretion rate can be extracted from \citet{2014ApJ...786...21P} and is given by
\begin{eqnarray}
\label{eq:Mdotenv}
 \dot{M}_{\rm gas} &= 0.00175 f^{-2} \left(\frac{\kappa_{\rm env}}{1{\rm cm}^2/{\rm g}}\right)^{-1} \left( \frac{\rho_{\rm c}}{5.5 {\rm g}/{\rm cm}^3} \right)^{-1/6} \left( \frac{M_{\rm c}}{{\rm M}_{\rm E}} \right)^{11/3} \nonumber \\ 
 &\left(\frac{M_{\rm env}}{0.1{\rm M}_{\rm E}}\right)^{-1} \left( \frac{T}{81 {\rm K}} \right)^{-0.5} \frac{{\rm M}_{\rm E}}{{\rm Myr}}
,\end{eqnarray}
where $f$ is a fudge factor to change the accretion rate in order to match numerical and analytical results, which is normally set to $f=0.2$ \citep{2014ApJ...786...21P}. The opacity in the planets envelope $\kappa_{\rm env}$ is generally very hard to determine because it depends on the grain sizes and distribution inside the planetary atmosphere. Here we use $\kappa_{\rm env} = 0.05{\rm cm}^2/{\rm g}$, which is very similar to the values used in the study by \citet{2008Icar..194..368M}. In Appendix~\ref{ap:kappa} we test the influence of different values of $\kappa_{\rm env}$ for gas accretion. For the density of the core, we assume $\rho_{\rm c}=5.5 {\rm g}/{\rm cm}^3$. This contraction phase ends as soon as $M_{\rm core}=M_{\rm env}$ and rapid gas accretion starts.

For rapid gas accretion ($M_{\rm core} < M_{\rm env}$), we follow \citet{2010MNRAS.405.1227M} directly. They calculated the gas accretion rate using 3D hydrodynamical simulations with nested grids. They find two different gas accretion branches, which are given as 
\begin{equation}
 \dot{M}_{\rm gas,low} = 0.83 \Omega_{\rm K} \Sigma_{\rm g} H^2 \left( \frac{r_{\rm H}}{H} \right)^{9/2}
\end{equation}
and
\begin{equation}
 \dot{M}_{\rm gas,high} = 0.14 \Omega_{\rm K} \Sigma_{\rm g} H^2 \ ,
\end{equation}
where the effective accretion rate is given by the minimum of these two accretion rates. The low branch is for low mass planets (with $(R_{\rm H}/h < 0.3$), while the high branch is for high mass planets ($(R_{\rm H}/h > 0.3$), and the effective accretion rate is given by the minimum value of both rates. Additionally, we limit the maximum accretion rate to $80\%$ of the disc's accretion rate onto the star, because gas can flow through the gap, even for high mass planets \citep{2006ApJ...641..526L}.

\subsection{Planet migration}
\label{subsec:migration}

The growing protoplanets inside the disc interact with the surrounding gas and migrate through it. The process of migration is substantially different between low mass planets that are still fully embedded in the disc (type-I migration) and high mass planets that open up a gap inside the disc (type-II migration). The migration rates of low mass planets can be obtained by 2D and 3D hydrodynamical simulations \citep{2008A&A...487L...9K, 2009A&A...506..971K, 2011A&A...536A..77B, 2014MNRAS.440..683L}. However, these simulations are very computationally intensive, so we use a prescribed formula to compute the torque acting on embedded planets \citep{2011MNRAS.410..293P}. The torque formula of \citet{2011MNRAS.410..293P} includes the effects of torque saturation and has been tested against 3D simulations in fully radiative discs \citep{2011A&A...536A..77B}, which find good agreement. Recent studies of \citet{Lega2015} tested the torque formula against numerical simulations in accreting discs including stellar and viscous heating and radiative cooling, as used in the disc model of \citet{2014arXiv1411.3255B}, and found very good agreement with respect to the zero-torque location in the disc, where planets would stop their inward migration. For low mass planets ($M_{\rm P}<5M_{\rm Earth}$), previous studies of \citet{2014MNRAS.440..683L} have shown a slight discrepancy with the torque formula of \citet{2011MNRAS.410..293P}, however these differences were found to be very small, even considering that very small planets migrate very slowly in the first place. Here we also assume that planets move only on circular orbits around the stars, because eccentricity and inclination is damped quite quickly by the gas disc \citep{2010A&A.523...A30, 2011A&A...530A..41B}.

We therefore use the torque formula of \citet{2011MNRAS.410..293P}, where the total torque $\Gamma_{\rm tot}$ acting on a planet is given as a composition of the Lindblad torque $\Gamma_{\rm L}$ and the corotation torque $\Gamma_{\rm C}$, 
\begin{equation}
 \Gamma_{\rm tot} = \Gamma_{\rm L} + \Gamma_{\rm C} \ .
\end{equation}
The Lindblad and corotation torques depend on the local radial gradients of surface density $\Sigma_{\rm g} \propto r^{-s}$, temperature $T \propto r^{-\beta}$, and entropy $S \propto r^{-\xi}$, with $\xi = \beta - (\gamma - 1.0) s$, where $\gamma=1.4$ is the adiabatic index. 

Very roughly said, for $\Sigma_{\rm g}$ gradients that are not too negative, a radially strong negative gradient in entropy, caused by a large negative gradient in temperature (large $\beta$), will lead to outward migration, while a shallow gradient in entropy will not lead to outward migration and planets migrate inwards. Therefore planets can migrate outwards in certain regions of the disc, where strong negative gradients in temperature can be found. Generally these regions of outward migration exist close to transitions in opacity, where $H/r$ drops \citep{2013A&A...549A.124B, 2014A&A...564A.135B, 2014arXiv1411.3255B}. However, as the disc evolves in time, these regions of outward migration also evolve in time, so that at the very late stages of the disc evolution ($\dot{M} <4 \times 10^{-9} M_\odot$/yr) only very small regions of outward migration exist that can only hold planets of up to $\approx 10 M_{\rm Earth}$. 

\begin{figure}
 \centering
 \includegraphics[scale=0.7]{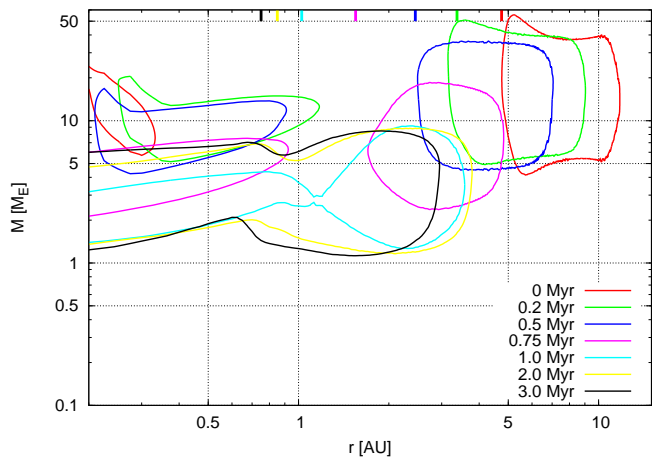}
 \caption{Regions for outward migration at different times for planets in type-I-migration using the torque expression of \citet{2011MNRAS.410..293P} for our standard disc structure with $Z_{\rm dust} = 0.5\%$. If the planet is located inside the solid lines, it will migrate outwards. If the planet is outside the solid lines, it will migrate inwards. The small ticks on the top of the plot correspond to the location of the ice line at the times indicated by the different colours.
   \label{fig:Migcont}
   }
\end{figure}

This is illustrated in Fig.~\ref{fig:Migcont} where we display the evolution of the regions of outward migration in time in our simulations. As the disc evolves in time, the strong negative gradient in entropy, which is caused by the negative gradient in temperature just outside the ice line ($r>r_{\rm ice}$) that can trigger outward migration, moves towards the star. The region of outward migration therefore moves towards the star, which finally makes the inner region of outward migration (caused by the silicate line) disappear after $0.75$ Myr. In the late stages outward migration is only possible in the inner parts of the disc for very low mass planets ($M_{\rm P} < 10 {\rm M}_{\rm E}$), where the regions of outward migration stay roughly constant, because the disc's temperature gradients do not evolve significantly at these stages of disc evolution any more \citep{2014arXiv1411.3255B}.

Planets that have reached their pebble isolation mass (eq.~\ref{eq:Misolation}) start to accrete gas (section~\ref{subsec:gasaccrete}) and grow even further until they finally open a gap inside the disc. A gap can be opened (with $\Sigma_{\rm Gap} < 0.1 \Sigma_{\rm g}$), when
\begin{equation}
\label{eq:gapopen}
 \mathcal{P} = \frac{3}{4} \frac{H}{r_{\rm H}} + \frac{50}{q \mathcal{R}} \leq 1 \ ,
\end{equation}
where $r_H$ is the Hill radius, $q=M_{\rm P} / M_\star$, and $\mathcal{R}$ the Reynolds number given by $\mathcal{R} = r_{\rm P}^2 \Omega_{\rm P} / \nu$ \citep{2006Icar..181..587C}. If the planet becomes massive enough to fulfil this criterion, it opens up a gap in the disc, and it migrates in the type-II regime. The gap-opening process splits the disc in two parts, which both repel the planet towards the centre of the gap, meaning that the migration time scale of the planet is the accretion time scale of the disc, $\tau_{\rm visc} = r_P^2 / \nu$. However, if the planet is much more massive than the gas outside the gap, it will slow down the viscous accretion. This happens if $M_{\rm P} > 4\pi \Sigma_{\rm g} r_{\rm P}^2$, which leads to the migration time scale of
\begin{equation}
\label{eq:typeII}
 \tau_{\rm II} = \tau_{\nu} \times \max \left(1 , \frac{M_{\rm P}}{4\pi \Sigma_{\rm g} r_{\rm P}^2} \right) \ ,
\end{equation}
resulting in slower inward migration for massive planets \citep{2013arXiv1312.4293B}.

Before the planet is massive enough to open up a deep gap inside the disc, there is still material inside the planets corotation region, which reduces the total negative torque acting on the planet and can in principle be so strong that the planet can migrate outwards \citep{2007MNRAS.377.1324C}. The depth of the gap is given in \citet{2007MNRAS.377.1324C} as
\begin{equation}
 f(\mathcal{P}) = \left\{
  \begin{array}{cc}
   \frac{\mathcal{P}-0.541}{4} &\quad \text{if} \quad \mathcal{P}<2.4646 \\
   1.0-\exp\left(-\frac{\mathcal{P}^{3/4}}{3}\right) &\quad \text{, otherwise.}
  \end{array}
  \right. \end{equation}
The factor $f(\mathcal{P})$ reduces the migration rate of the planet, when a partial gap is opened in the disc, because the migration rate depends directly on the gas surface density $\Sigma_{\rm g}$. We multiply our migration rate directly by $f(\mathcal{P})$ to reduce the migration rate by the partial opening of the gap. This reduction of the migration rate is very crucial, because planets that open a partial gap in the disc generally have several $10$s of Earth masses and migrate very fast through the disc, because they are still in the type-I migration regime.

In additional, we use a linear smoothing function for the transition between planets that open partial gaps inside the disc (that migrate with the reduced type-I speed by the factor $f(\mathcal{P})$) and planets that migrate with type-II, because even the reduced type-I migration rate (if the gap is fully opened with $\mathcal{P}<1$) is different from the nominal type-II rate.

A new study by \citet{2015Natur.520...63B} shows that low mass planets ($M_{\rm P}<5$ ${\rm M}_{\rm E}$) that accrete very fast can actually migrate outwards instead of the inward type-I-migration. However, we find that this effect is not that important for planets that grow via pebble accretion, because the growth is so fast that $5$ ${\rm M}_{\rm E}$ is reached in a very short time, which also limits the time the planet actually migrates until it reaches $5$ ${\rm M}_{\rm E}$. This is discussed in Appendix~\ref{ap:heating}.

The corotation torque arises from material that executes horseshoe U-turns relative to the planet, where most of this material is trapped in the planet's horseshoe region. But, if the planet migrates with respect to the disc, material outside the horseshoe region will execute a unique horseshoe U-turn relative to the planet, which can alter its migration speed. This becomes important, in particular, when the planet starts to carve a gap around its orbit. This can lead to runaway type-III migration, if the coorbital mass deficit $\delta M$ is greater than the mass of the planet $M_{\rm P}$, which can significantly change the semi-major axis of the planet in just a few orbits \citep{2003ApJ...588..494M}. The co-orbital mass deficit is defined as the mass that the planet pushed away from its orbit compared to the unperturbed disc structure as it starts to open up a gap. Unfortunately, there are no prescriptions to model this migration analytically, but we nevertheless test whether a growing planet might be subject to runaway type-III migration
during our simulations.

Migrating planets also experience dynamical torques that are proportional to the migration rate and depend on the background vortensity\footnote{Vortensity is defined as the ratio of vorticity and surface density.} gradient \citep{2014MNRAS.444.2031P}. These dynamical torques can have either positive or negative feedback on the migration, depending on whether the planet migrates with or against the direction of the static corotation torque. The effects of these dynamical corotation torques can be profound because they can slow down inward migration significantly, and outward migration can proceed beyond the zero-torque lines in discs that are massive and have a low viscosity. An approximate estimate of whether dynamical corotation torques play a role depends on the disc's viscosity and mass, as well as on the planet's mass \citet{2014MNRAS.444.2031P}. In particular, when
\begin{equation}
 \frac{\nu_{\rm P,0}}{r_{\rm P,0}^2 \Omega_{\rm P,0}} < \frac{16}{3\pi} \left( \frac{3}{2} - s \right) \frac{\Gamma_{\rm tot}/\Gamma_{\rm 0} q_{\rm d}^2 \bar{x}_{\rm s}^3}{(H/r)^2}
,\end{equation}
migration due to the dynamical corotation torques will be important. Here the subscripts $P$ and $0$ indicate the initial location of the planet, $s$ is the radial gradient of the surface density profile, $\Gamma_{\rm 0}$ the normalisation of the torque, $q_{\rm d}=\pi r_{\rm P,0}^2 \Sigma_{\rm P,0} / M_\star$ the disc mass, and $\bar{x}_{\rm s} = \sqrt{q/h}$, where $q=M_{\rm P}/M_\star$. Unfortunately this effect is not quantified further, so that we just indicate when the dynamical corotation torques might become important, but our simulations do not evolve with them. This effect and the effect of type-III migration are both discussed in Appendix~\ref{ap:typeIII}.

As soon as the planet reaches an inner edge of $0.1$~AU, we not only stop migration, but also the total simulation. That close to the central star, stellar tides can become important and influence the evolution of the planet, which we do not take into account.

\section{Formation of planets}
\label{sec:formation}

In this section we explore the growth of planetesimals that are inserted at a given initial time $t_{\rm 0}$ and a given initial distance $r_{\rm 0}$ into the disc. The initial time $t_{\rm 0}$ is important, because we keep the lifetime of the disc at $3$ Myr, meaning that planets that are inserted into the disc, for example at $t_{\rm 0}=2$ Myr, will only experience $1$ Myr of evolution. Additionally, the structure of the protoplanetary disc is different at different ages \citep{2014arXiv1411.3255B}. The initial distance $r_{\rm 0}$ determines where the planetary seed is placed. This strongly influences the initial growth of the planet, because the surface density $\Sigma_{\rm g}$ is lower at greater orbital distances, which means that the pebble surface density $\Sigma_{\rm peb}$ is also lower, indicating a longer growth time of the core. After the planets have reached their pebble isolation mass, the contraction of the gaseous envelope begins until $M_{\rm c} < M_{\rm env}$. At this point, runaway gas accretion starts.

We focus here on discs that have a total lifetime of $3$ Myr. A longer disc lifetime ($5$ Myr) does not affect our results qualitatively, but simply pushes the preferred planet formation time out to $\sim 3$~Myr (see Appendix~\ref{ap:decay}). Additionally, in this section we assume that the opacity in the envelope is fixed to $\kappa_{\rm env} = 0.05$ cm$^2/$g. Different opacities of the envelope are explored in Appendix~\ref{ap:kappa}.

In Table~\ref{tab:classes} we define the different planetary categories used in this work. Our definition of different planetary categories is simply a function of planetary mass. Only the subcategories are a function of the final orbital distance. The definition of ice giants is slightly different than in \citet{2014arXiv1408.6087L}, where ice giants are required to not reach pebble isolation mass. This change in the definition is related to the slow contraction of the gaseous envelope (eq.~\ref{eq:Mdotenv}), which allows for $M_{\rm c}>M_{\rm env}$ for a long time after reaching pebble isolation mass.

{%
\begin{table*}
 \centering
 \begin{tabular}{ccc}
 \hline \hline
 \textbf{Planet category} & \textbf{Planetary mass} & \textbf{Orbital distance} \\\hline
 {Ice planet} & $0.1 {\rm M}_{\rm E} < M_{\rm P} < 2{\rm M}_{\rm E}$ & - \\
 {Ice giant} & $M_{\rm P} \geq 2{\rm M}_{\rm E}$ and $M_{\rm c} \geq M_{\rm env}$ & - \\
 {Cold gas giant} & $M_{\rm c} < M_{\rm env}$ & $r_{\rm f} > 1.0$~AU  \\
 {Warm gas giant} & $M_{\rm c} < M_{\rm env}$ & $0.1$~AU $<r_{\rm f}<1.0$~AU  \\
 {Hot gas giant} & $M_{\rm c} < M_{\rm env}$ & $r_{\rm f} < 0.1$~AU  \\
 \hline
 \end{tabular}
 \caption{Definition of planets used in this work. We distinguish the main classes of planets (gas giants, ice giants, and ice planets) only by mass and mass ratio between core and envelope. The subcategories for gas giants are defined through their orbital distances, which we do not apply for ice giants and ice planets.
 \label{tab:classes}
 }
\end{table*}
}%

\subsection{Single evolution track}

\begin{figure}
 \centering
 \includegraphics[scale=0.6]{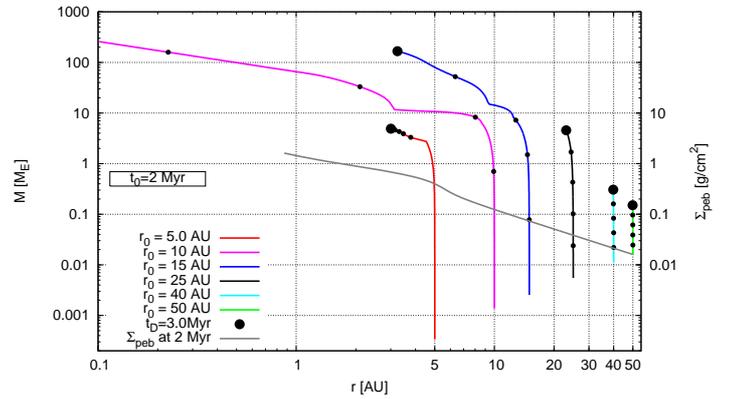}
 \caption{Growth tracks of planets that accrete pebbles in an evolving protoplanetary disc. The big black circular symbols indicate the final mass and position of the planets at $t_{\rm f}=3$ Myr, meaning that the planets experience $1$ Myr of evolution. The small black dots indicate $2.2$, $2.4$, $2.6,$ and $2.8$ Myr. In the example of the purple line ($r_0=10$~AU), the planet reaches $0.1$~AU before the final disc lifetime, and we stop the simulation. The different coloured lines indicate different planetary evolutions, which correspond to different types of planets, where we have an ice giant, a hot gas giant, a cold gas giant, one more ice giant, and two ice planets starting from inside out (see text). The grey line indicates the surface density of pebbles in the disc at $2$ Myr and ends at the ice line slightly interior to $1$~AU, where the pebble surface density changes. .
   \label{fig:Envhockey}
   }
\end{figure}

In this section we follow the evolution tracks of planets in the disc to clarify the different outcomes of planetary evolution. We use a metallicity of $Z=1.0\%$ in pebbles that can be accreted onto the initial seed masses. When we include the $0.5\%$ of metals in dust grains from the disc structure, our total metallicity is therefore roughly the solar value. We chose different initial starting locations $r_0$ of the planets, and all planets start at the initial time $t_{\rm 0}=2$ Myr, which means that the planets will evolve for $1$ Myr, because our total disc lifetime is $3$ Myr. In Fig.~\ref{fig:Envhockey} the evolution tracks of the planets and the pebble surface density $\Sigma_{\rm peb}$ at $2$ Myr are shown. The surface density of pebbles $\Sigma_{\rm peb}$ is calculated through eq.~\ref{eq:SigmaPeb}, which depends on $\sqrt{\Sigma_{\rm g}}$. Therefore the bumps in $\Sigma_{\rm g}$ caused by the transitions in opacity  \citep{2014arXiv1411.3255B} translate into bumps in the pebble surface density. As the gas surface density decreases and evolves in time, so does the pebble surface density. The starting mass of the planets is different at different locations, because it is set by the pebble transition mass, which depends on the disc's aspect ratio (eq.~\ref{eq:Mtrans}). 

Planets that manage to reach their pebble isolation mass contract their envelope until $M_{\rm c} < M_{\rm env}$ and the planets can start runaway gas accretion. During the whole growth process the planet migrates through the disc. The planets have been inserted into the disc at an disc age of $t_0=2$ Myr and are evolved until $t_{\rm f} = 3$ Myr. Planets that cross the ice line have already reached pebble isolation mass, so they do not accrete pebbles any more, making it unnecessary to model a transition in pebble size and pebble surface density at the ice line.

The planetary seed implanted at $50$~AU (green line in  Fig.~\ref{fig:Envhockey}) accretes pebbles very slowly, because the pebble column density is very low at those orbital distances. At $3$ Myr, the planet has just reached $\approx 0.2$ ${\rm M}_{\rm E}$ and is very far away from reaching pebble isolation mass. Because of its low mass, it also migrates only a very short distance. This is a typical example of an {\it \emph{ice planet}} in our definition.

The planet starting a bit farther inside at $r_0=40$~AU (light blue line in Fig.~\ref{fig:Envhockey}) grows also very slowly, because the surface density in pebbles and the Stokes number ($\tau_{\rm f} < 0.1$) is low in the outer disc. It therefore never reaches pebble isolation mass and stays low, with a mass of only a few tenths of an Earth mass. Because the planet stays quite small, it only migrates a few AU inwards. This is an example of a slightly more massive ice planet in our definition, because it forms in the cold outer parts of the disc and is less massive than $2$ ${\rm M}_{\rm E}$. The planets starting at $r_0=40$ and $r_0=50$~AU always accrete in the 3D scheme, because (i) the particle scale height is large in the outer disc and (ii) $\tau_{\rm f}$ is very small, which explains their low growth rates.

The planetary seed starting at $25$~AU (black line in Fig.~\ref{fig:Envhockey}) accretes pebbles at a faster rate (higher $\Sigma_{\rm peb}$, larger $\tau_{\rm f}$), but it misses reaching pebble isolation mass after $1$ Myr of evolution. Because it is more massive, it migrates farther in the disc and ends up at $\sim 22$ AU. Its mass ends up at about $5~{\rm M}_{\rm E}$, where the mass of the core is much more massive than the mass of the envelope, making it an {\it \emph{ice giant}} according to our definition.

Starting a planetary seed at $15$~AU (dark blue line in Fig.~\ref{fig:Envhockey}) reveals a new growth path. After the planet has reached pebble isolation mass, its envelope contracts and the planet starts a runaway gas accretion process, making it a {\it \emph{gas giant}} planet. During its growth the planet migrates inwards from $15$~AU down to $\sim 3$~AU. The main migration happens when the planet is undergoing fast inward type-I migration, before it is massive enough to open a gap in the disc (at $\sim 120$ ${\rm M}_{\rm E}$). This fast inward migration before gap opening indicates that the formation of gas giants requires a formation much farther out in the disc than their final orbital position would indicate; {\it \emph{in situ}} formation of gas giants is impossible when including full planetary migration rates.

The planetary seed starting at $10$~AU (purple line in Fig.~\ref{fig:Envhockey}) follows a similar evolution to the one starting at $15$ AU, but with two notable exceptions. After the planet has reached pebble isolation mass, it migrates inwards very rapidly to $3$ AU, without growing too much. This fast inward migration is caused by the disc structure, where the region between $3$ and $8$~AU has a very shallow (and even inverted) radial temperature gradient \citep{2014arXiv1411.3255B}, which causes a strong negative total torque acting on the planet driving fast inward migration. The planet is then in a region of very slow inward migration (just a bit too massive to be caught in the region of outward migration, Fig.~\ref{fig:Migcont}), where it migrates very slowly and starts to rapidly accrete gas. During this accretion process, the planet migrates further into the inner regions of the disc. In fact, the planet migrates so fast that it reaches the inner edge of the disc at $0.1$ AU, before the end of the lifetime of the disc is reached. (That is also why there is no big black circle in Fig.~\ref{fig:Envhockey} for this planet.) As soon as the planet reaches $0.1$ AU, we stop the simulation. The planet has become a {\it \emph{hot gas giant}}.

If the planet starts in the very inner regions of the disc at $5$~AU (red line in Fig.~\ref{fig:Envhockey}), its isolation mass is very low, because $H/r$ is very small in the inner parts of the disc at an evolution stage of $2$ Myr. It therefore accumulates only $2.5$ Earth masses of solids. A low core mass then leads to a very long contraction time of the envelope (eq.~\ref{eq:Mdotenv}), resulting in a total planet mass of only a few Earth masses when the disc reaches an age of $3$ Myr. During the evolution, the planet is small enough to be caught at the zero-migration distance for most of its evolution, meaning that the planet follows the zero-migration distance as the disc accretes onto the star (see Fig.~\ref{fig:Migcont}). This planet is also classified as an \emph{{\it \emph{ice giant}}} because it formed at $r_{\rm P}>r_{\rm ice}$ (Fig.~\ref{fig:Migcont}).

We would like to point out here that definitions of planets in this work is only related to the planetary mass and the mass ratio between the planetary core and envelope. The final orbital distance only plays a role in the subcategories of planets, so that gas giants very close to the central star ($r_{\rm f} < 0.1$~AU) are called {\it \emph{hot}} gas giants (Table~\ref{tab:classes}).

\subsection{Variation in the initial orbital position}

We can now expand Fig.~\ref{fig:Envhockey} over the whole radial domain of the disc, but keep the initial time when we insert the planet fixed at $t_{\rm 0}=2$ Myr. At each orbital distance $3$~AU $<r_{\rm 0}<50$~AU, a single planet is put into the disc and evolved independently. The nominal lifetime of the disc is $3$ Myr, so the planets evolve for $1$ Myr in the disc.

\begin{figure}
 \centering
 \includegraphics[scale=0.7]{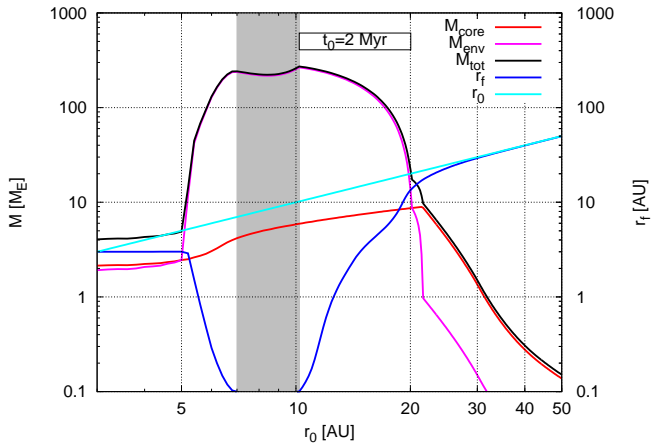}
 \caption{Final planetary core mass $M_{\rm core}$ and corresponding envelope mass $M_{\rm env}$, as well as total mass $M_{\rm tot} = M_{\rm c} + M_{\rm env}$ and final orbital distance $r_{\rm f}$ of planets in the evolving disc at a disc dissipation time of $3$ Myr. The light blue line represents the initial orbital distance $r_0$. Planets inside the grey area are subject to very strong migration, so they end up in the inner parts of the disc with $r_{\rm f}<0.1$~AU and become hot gas giants. The divide between ice and gas giants can be seen easily at $\sim 22$~AU, where $M_{\rm c}<M_{\rm env}$. All data lines for one $r_0$ location in this plot give the final result of an individual growth track as in Fig.~\ref{fig:Envhockey}. 
   \label{fig:Envelope2Myr}
   }
\end{figure}

In Fig.~\ref{fig:Envelope2Myr} we present the final planetary core mass $M_{\rm c}$ and envelope mass $M_{\rm env}$, as well as total planetary mass $M_{\rm tot}$ with respect to the initial orbital position $r_0$ and the final orbital distance $r_{\rm f}$ of planets that evolved in the disc for $1$ Myr after insertion at $t_{\rm 0}=2$ Myr. Planets inside the grey area are within $r_{\rm f}<0.1$~AU to the host star, and their evolution is stopped before $1$ Myr of evolution is reached. The planetary masses and final positions are displayed as a function of their initial orbital distance $r_{\rm 0}$, which illustrates what influence the initial orbital distance of the planet has on the final properties of the planet. For example, when looking at $r_0=15$~AU, the total planetary mass $M_{\rm tot}$ and final orbital distance $r_{\rm f}$  shown correspond to the location of the big black circle of the $r_0=15$~AU planet in Fig.~\ref{fig:Envhockey}.

Planets forming outside of $r_{\rm 0}>22$~AU do not reach pebble isolation mass, because the pebble density in the outer disc is very low and growth time too long. But, these planets can still have a few Earth masses, making them ice giants by our definition, because $M_{\rm c} > M_{\rm env}$ and $M_{\rm P} > 2 {\rm M}_{\rm E}$. However, planets starting with $r_0>28$~AU grow only very little, because of a small $\tau_{\rm f}$ and a larger scale height, which only allows planets to grow with the 3D accretion mechanism. These planets then have $M_{\rm P} < 2{\rm M}_{\rm E}$ and are ice planets instead.

Just inside ($20$~AU $< r_{\rm 0}<22$~AU), the planets have reached pebble isolation mass, but did not reach runaway gas accretion, because $M_{\rm c}>M_{\rm env}$, making these planets ice giants. Planets that form farther inside with $20$~AU $>r_{\rm 0}>13$~AU become gas giants that stay in the outer disc with $r_{\rm f}>1$~AU. This indicates that there is a very broad range of radial extent that allows for the formation of gas giants, because their cores still form quickly enough via pebble accretion that enough time is left to accrete a gaseous envelope before the disc dissipates. With increasing $r_0$, the mass of the core increases, which is caused by the flaring structure of the disc at that evolutionary stage \citep{2014arXiv1411.3255B}, which increases the pebble isolation mass (eq.~\ref{eq:Misolation}).

Planets that form within $r_{\rm 0} <13$~AU and outside of $6$~AU all end up within $r_{\rm f} <1$~AU and are gas giants. The reason for their strong inward migration lies in the efficient growth of the planetary core via pebbles. Because the core grows very quickly, the planets have a longer time to migrate faster compared than cores that grow more slowly, because the type-I migration speed is proportional to the planetary mass. The planets then accrete gas to become gas giants and open up a gap in the disc that slows their migration (type-II migration). However, planets that form in a region of $7$~AU$<r_0<10.5$~AU grow too quickly and migrate too fast to stay outside of $0.1$~AU, indicating that planets that form in this part of the disc will end up as hot gas giants. When these planets reach $r<0.1$~AU, the total evolution of the planet is stopped, explaining the kink in the gas mass of planets inside the grey area in Fig.~\ref{fig:Envelope2Myr}.

Planets that form in the inner regions of the disc $r_0<7$~AU only grow a very small planetary core even though enough pebbles are available, because the pebble isolation mass is low. The low core mass then prevents a fast accretion of the envelope, so that a long time is needed for the contraction of the envelope. This prevents planets that formed with $r_0<5$~AU from accreting a massive gaseous envelope, and these planets stay with $M_{\rm c}>M_{\rm env}$, indicating that these planets are ice giants. These planets are also caught in a region of outward migration, letting them stay a few AU from the central star (Fig.~\ref{fig:Migcont}). The planets that are within $5$~AU and $7$~AU have a slightly more massive core, indicating a slightly faster contraction phase of the envelope, so these planets can rapidly accrete gas and thus form gas giants that have then outgrown the region of zero migration and migrate towards the star.

Our model predicts the formation of different types of planets by their initial formation location $r_{\rm 0}$. Additionally, our model predicts that gas giant planets that have $r_{\rm f}>1$~AU form in the outer regions of the disc ($r_{\rm 0}>13$~AU) and thus do not form {\it \emph{in situ}}. This is a big contrast to the study of \citet{2014arXiv1407.6011C}, where the cores of giant planets are built locally at a few AU by the accretion of planetesimals and planetary embryos. When the planetary cores then become massive enough,  they can be caught in a region of outward migration (Fig.~\ref{fig:Migcont}) and then accrete gas to form a gas giant. However, \citet{2014arXiv1407.6011C} did not include gas accretion and therefore did not observe the inward migration of giant planets in type-II migration, which can bring them very close to the host star, especially when they form just a few AU away from it. In our study we overcome this problem because pebble accretion is very efficient in the outer disc and thus allows the formation of planets at large orbital distances, which allows them to stay far away from the host star even after their inward migration.

\subsection{Variation in the initial time and position}

\begin{figure*}
 \centering
 \includegraphics[width=17cm]{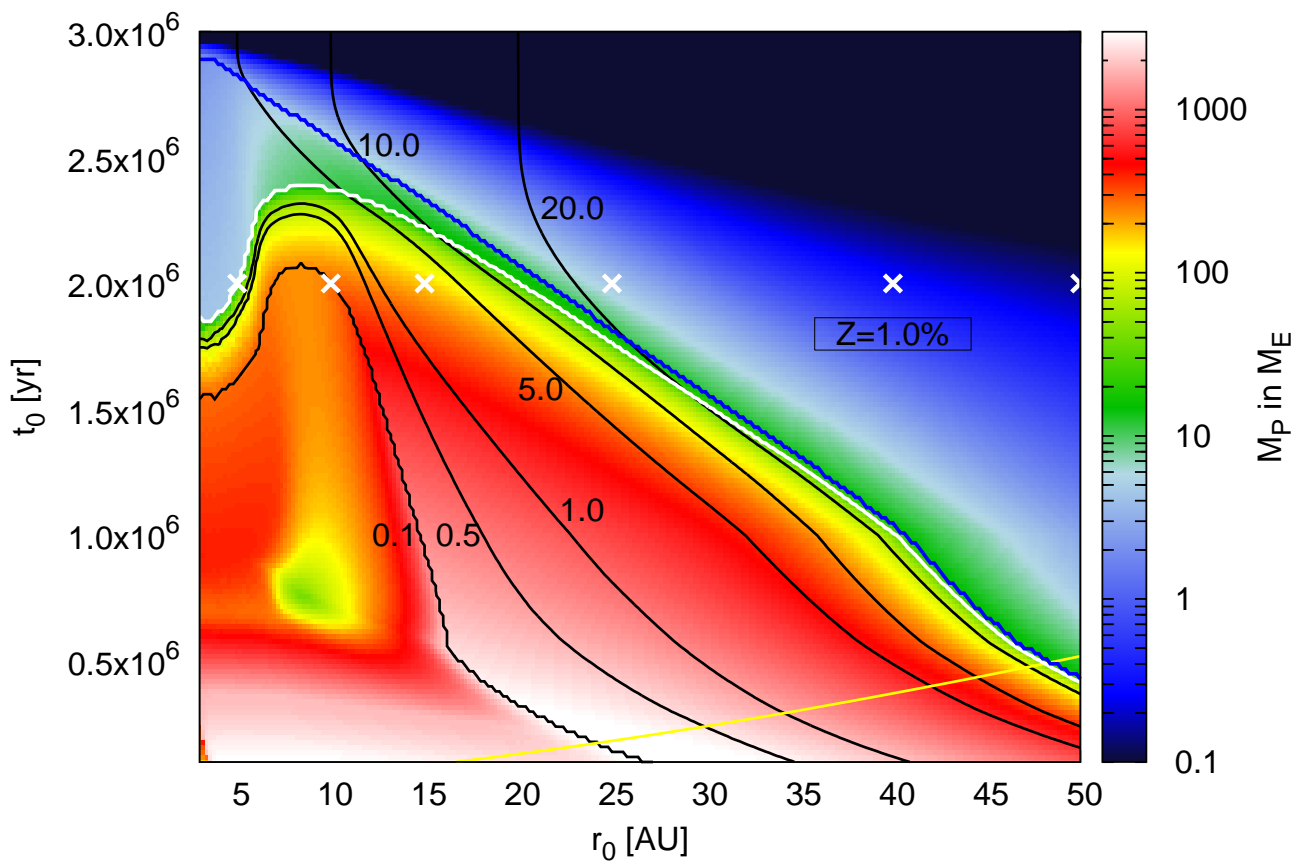}
 \caption{Final masses of planets (total mass $M_{\rm P}=M_{\rm c}+M_{\rm env}$) as a function of formation distance $r_{\rm 0}$ and formation time $t_{\rm 0}$ in the disc. Planets that are below the dark blue line have reached pebble isolation mass and can accrete gas. All planets that are below the white line have $M_{\rm c}<M_{\rm env}$, indicating that they have undergone runaway gas accretion. The yellow line marks the pebble production line; planets below the yellow line cannot have formed by pebble accretion, because the pebbles have not yet formed at the insertion time of the planet. The black lines indicate the final orbital distance $r_{\rm f}$ of the planet. Each point in $(r_{\rm 0},t_{\rm 0})$ corresponds to the final mass of one individual evolution track as shown in Fig.~\ref{fig:Envhockey}. The white crosses indicate the initial orbital positions and times for the planetary evolution tracks shown in Fig.~\ref{fig:Envhockey}.
   \label{fig:Z008kappa}
   }
\end{figure*}

We now expand Fig.~\ref{fig:Envelope2Myr} in the dimension of initial time $t_{\rm 0}$ when we place the planetary seed inside the disc. We start planets in our disc from $t_{\rm 0}=100$ kyr up to the end of the disc's lifetime of $t_{\rm D}=3$ Myr, as well as with $3$~AU$<r_{\rm 0}<50$~AU. Figure~\ref{fig:Z008kappa} presents the final planetary mass as a function of initial radius $r_0$ and initial time $t_0$. Each point in this figure corresponds to the final mass of a growth simulation as in Fig.~\ref{fig:Envhockey}. All line cuts for a fixed initial time $t_{\rm 0}$ with all $3$~AU$<r_{\rm 0}<50$~AU correspond to a plot similar to Fig.~\ref{fig:Envelope2Myr}. In particular, Fig.~\ref{fig:Envelope2Myr} represents a cut at $t_{\rm 0}=2$ Myr of Fig.~\ref{fig:Z008kappa}.

In Fig.~\ref{fig:Z008kappa} the region $r_{\rm 0}$ and formation time $t_{\rm 0}$ in combination with the final orbital distance $r_{\rm f}$ can be interpreted as a map for the formation of different types of planets. Within $r_{\rm f}<0.1$~AU, we find hot gas giants. A late formation time $t_{\rm 0}$ prolongs the formation time of the core, because fewer pebbles are available, and reduces the mass of the core at pebble isolation, because $H/r$ drops in time. This means that the planet spends more time contracting its envelope, meaning that it can not accrete as much gas. Moreover, the late formation time reduces the time the planet migrates inwards in the disc, letting it stay farther out in the disc.

In the band of $0.1$au$<r_{\rm f}<15.0$~AU, we find warm and cold gas giant planets. The formation of these planets can also occur very easily until $t_{\rm 0}\approx 2$ Myr, because the pebble accretion rate is high even when the planets form at a large $r_{\rm 0}$. This formation can then compensate for inward type-I migration through the first phases of gas accretion before the planets open up a gap in the disc and transitions into slow type-II migration. It also provides enough time to accrete a massive gaseous envelope after the long contraction phase. 

The planets around $t_0 \approx 1$ Myr and $r_0\approx 10$~AU start in the flaring part of the disc, meaning that they have a have higher pebble isolation mass than the planets with smaller $r_0$, so they can accrete their envelope faster and become more massive. At the same time, these planets are too massive to spend time in the region of outward migration, because they are already too massive when they reach this region (Fig.~\ref{fig:Migcont}), which results in a continuous inward migration of these planets. They therefore reach $0.1$~AU before they become more massive than $\sim 300$ Earth masses.

Forming at a later time or an even larger $r_{\rm 0}$ results in smaller gas giants that are more in the mass regime of Saturn. Saturn-mass planets form in a distinct band below the white line in Fig.~\ref{fig:Z008kappa}, where everything below the white line has $M_{\rm c}<M_{\rm env}$, and runaway gas accretion has set in. All planets between the blue and the white lines in Fig.~\ref{fig:Z008kappa} have reached their pebble isolation mass and started to contract their envelope, but have $M_{\rm c}>M_{\rm env}$. In the inner disc these planets are only a few Earth masses and qualify as ice giants (light blue background colour in Fig.~\ref{fig:Z008kappa} between the blue and white lines), while in the outer disc the planets become more massive ($M_{\rm P} > 10 {\rm M}_{\rm E}$) and are larger ice giants. The implications of our model regarding the formation of the ice giants in our own solar system are discussed in more detail in section~\ref{sec:solarsystem}.

In the late stages of the disc ($t>2$ Myr), the region of outward migration is located at $\sim 3$~AU and can hold planets of up to a few Earth masses (Fig.~\ref{fig:Migcont}). This means that ice giants stop their inward migration there, explaining the pile-up of ice giants in this region of parameter space.

We want to note that Fig.~\ref{fig:Z008kappa} does {\it \emph{not}} give a percentage of what kind of planets should exist around other stars. Figure~\ref{fig:Z008kappa} instead shows what kind of planet would form if an initial seed planetesimal with $M_{\rm trans}$ (eq.~\ref{eq:Mtrans}) was placed in the disc at $r_{\rm 0}$ and $t_{\rm 0}$ for a disc that lives, in total $3$ Myr, around a solar type star. We can learn from Fig.~\ref{fig:Z008kappa} that gas giants are more abundant if an early formation scenario is invoked and that hot gas giants have to form early and fairly close to the central star, while cold gas giants can form at later times and further out in the disc. {\it In situ} formation of gas giants is not possible. For later formation times ($t_0>2$ Myr), the final mass of the planet becomes lower, because the isolation mass becomes lower as $H/r$ drops, resulting in smaller planetary cores. A smaller planetary core then prolongs the contraction time of the gaseous envelope, allowing the formation of planets that have $M_{\rm c}>M_{\rm env}$ at $2$ Myr, because the planets have a shorter total evolution time. Considering that the most common observed exoplanets are small ($M_{\rm P} < 10 {\rm M}_{\rm E}$), a later formation time for planets is favoured by our planet formation scenario.

These results emphasise the importance of an evolving disc structure that is not a simple power law. We discuss planet formation via pebble accretion in the MMSN in Appendix~\ref{ap:MMSN}, where we show that the formation of different planetary types is dramatically different when adopting too simplistic a disc model.

\subsection{Influence of pebble surface density}

The metallicity in pebbles $Z$ has a strong influence on the outcome of planetary systems \citep{2014arXiv1408.6094L}. A higher metallicity in pebbles will allow for faster growth of the planetary core, resulting in a larger core. However, the pebble isolation mass (eq.~\ref{eq:Misolation}) reduces in time as the disc evolves and planets form, keeping the final core mass from being twice as high for discs with twice the metallicity in pebbles. Lower metallicity in pebbles slows down the growth of the core of the planet, resulting in lower core mass, because the isolation mass drops in time owing to $H/r$ decreasing in the longer time that is needed to build the core. This change in the mass of the core will then influence the gas accretion onto the planet, because a larger planetary core can contract its envelope in a shorter time than a smaller planetary core. The mass of the cores of these planets in a $Z=1.5\%$ disc is $\sim 20-30 {\rm M}_{\rm E}$, which is about a factor of three to four too low to explain the amount in heavy elements for Corot-13 b, 14 b, 17 b, and 23 b, which have solid cores of around $100 {\rm M}_{\rm E}$ \citep{2013Icar..226.1625M}.

\begin{figure}
 \centering
 \includegraphics[scale=0.7]{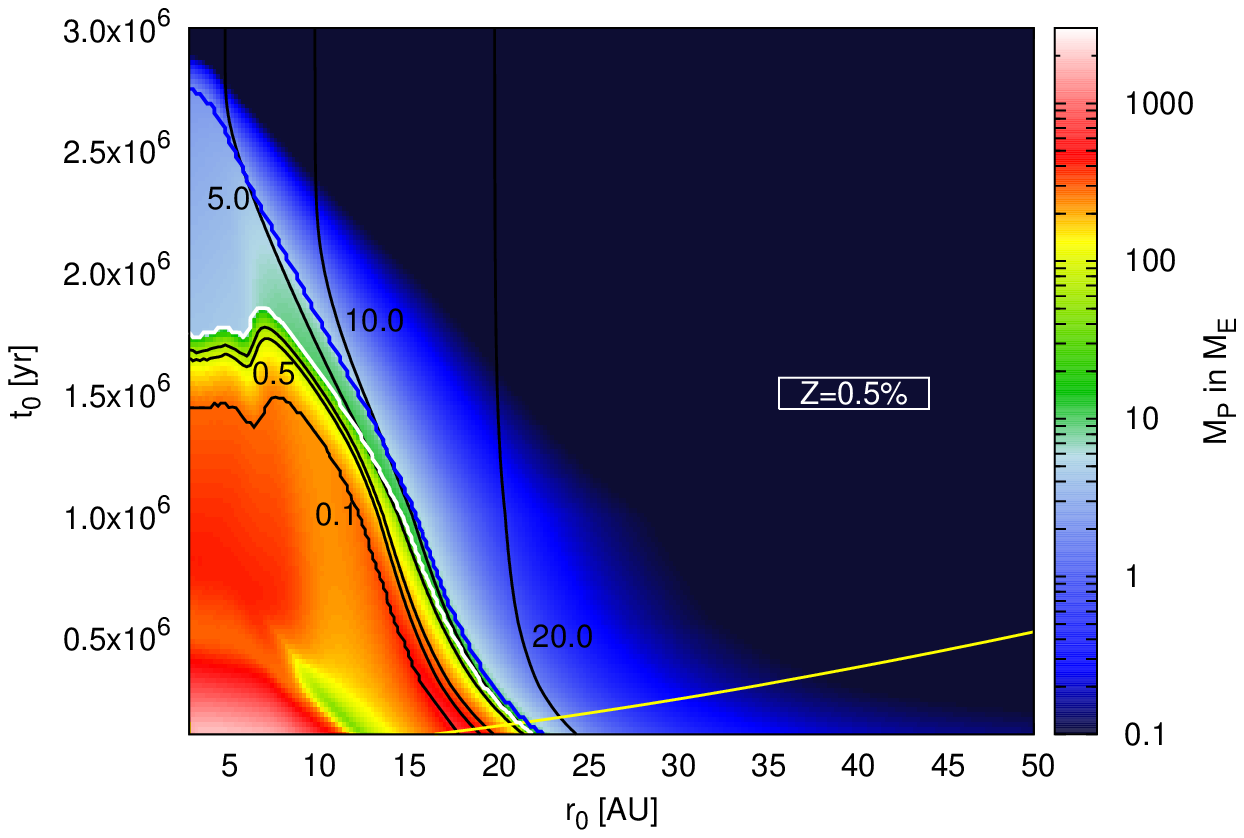}
 \includegraphics[scale=0.7]{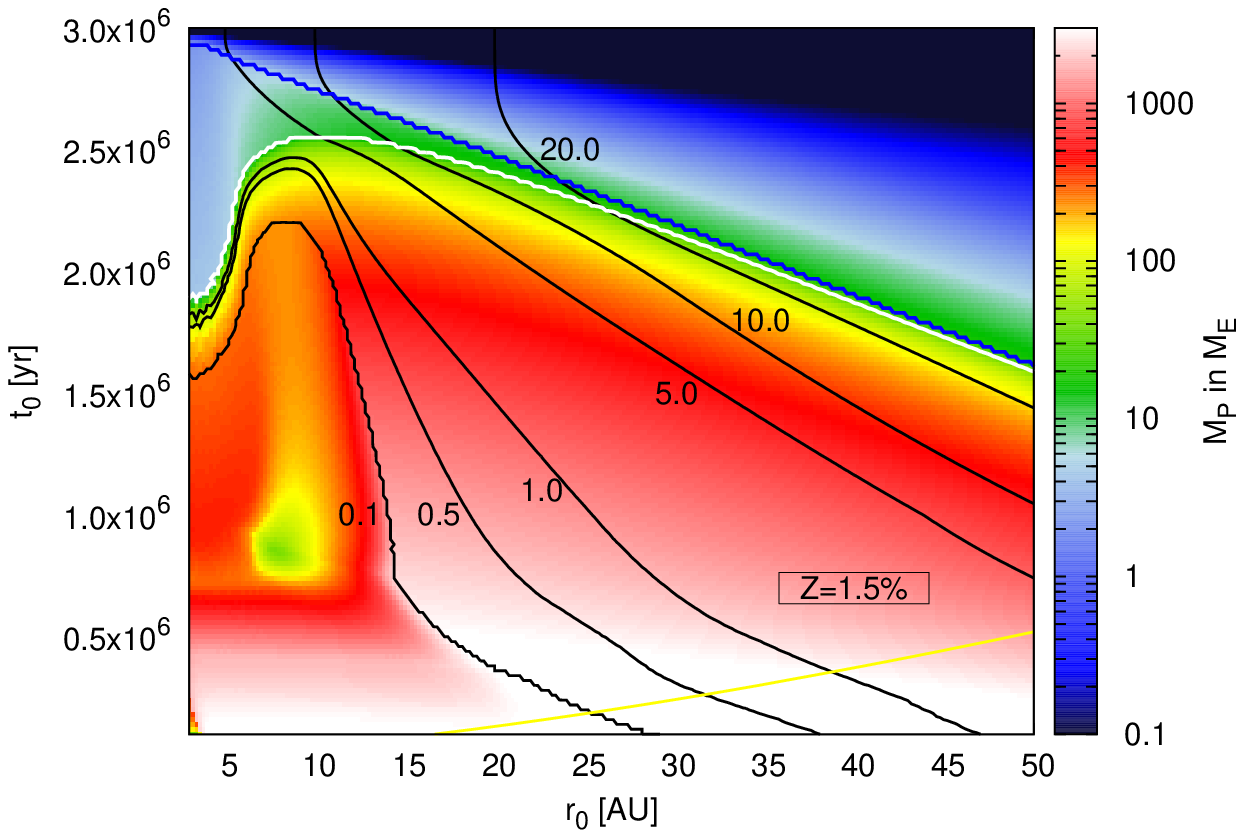}
 \caption{Final masses of planets ($M_{\rm P} = M_{\rm c}+M_{\rm env}$) as a function of initial radius $r_{\rm 0}$ and initial time $t_{\rm 0}$ in the disc. The top plot features a metallicity of $Z=0.5\%$ in pebbles, the bottom plot a metallicity of $Z=1.5\%$. The different lines inside the plots have the same meaning as in Fig.~\ref{fig:Z008kappa}.
   \label{fig:Zenvelope}
   }
\end{figure}

In Fig.~\ref{fig:Zenvelope} we display the final planetary masses for $Z=0.5\%$ (top, half the nominal metallicity in pebbles) and for $Z=1.5\%$ (bottom, $3/2$ times the nominal metallicity in pebbles). In general, we still find the same planetary classes in the $r_{\rm 0}-t_{\rm 0}$ plane, but its distribution is different. 

In the low-metallicity case, the maximum $t_0$ for forming a gas giant planet is reduced compared to the $Z=1.0\%$ case, so that for $t_0>2$ Myr, no gas giants can form any more. The region of parameter space that results in hot gas giants ($r_{\rm f}<0.1$~AU) extends to larger $r_0$ for a given $t_0$. This is caused by the fact that the planets in a disc with $Z=0.5\%$ have smaller planetary cores, which results in a longer contraction time for the envelope. The planet therefore spends a longer time in the fast inward type-I migration regime, before it grows big enough via gas accretion to open a gap in the disc, which in total results in a smaller $r_{\rm f}$. This is another indication that $2$ Myr is a reasonable seed formation time, because otherwise we observe more hot gas giants around lower metallicity stars.

This effect not only influences the parameter space that harbours hot gas giants, but it also influences all the parameter space where gas giants are formed. As a result, the region of parameter space with gas giants with $0.1$~AU $<r_{\rm f}<5$~AU is much smaller in the $Z=0.5\%$ case than in the $Z=1.0\%$ case, indicating that the gas giants should be rarer in discs with lower metallicity. Likewise, the region of parameter space for forming ice giants is slightly enlarged in the $Z=0.5\%$ case, consistent with observations \citep{2014Natur.509..593B}.

In the case with higher metallicity, $Z=1.5\%$, the parameter space for gas giants with $r_{\rm f}<0.1$~AU is a bit smaller than in the $Z=1.0\%$ case. Higher metallicity helps to keep planets outside of $0.1$~AU because the formation time of the core is shorter and the core is larger, also
making the phase of envelope contraction shorter, which in turn allows rapid gas accretion at an earlier stage. This then results in a planet being able to open a gap soon and migrate slower in type-II migration speed, allowing for larger $r_{\rm f}$ in the end. Along with that, the more massive planets slow down their type-II migration rate because of the feedback with the disc (Eq.~\ref{eq:typeII}). However, for later initial times $t_{\rm 0}> 1$ Myr, this effect does not matter that much compared to the $Z=1.0\%$ disc, because the planets do not reach such high masses that the reduction of the type-II migration speed plays that much of a role. Another aspect is that the formation of gas giants is now possible up to $t_{\rm 0}=2.5$ Myr at nearly all initial orbital distances $r_{\rm 0}$, except for the inner regions of the disc.

On the other hand, the high metallicity in pebbles reduces the $r_{\rm 0}-t_{\rm 0}$ parameter space for the formation of ice giants slightly compared to simulations with low $Z$. A higher pebble accretion rate allows for efficient formation of planetary cores, even at late stages (high $t_{\rm 0}$) of the disc evolution, resulting in a more massive core than in the $Z=1.0\%$ case, in turn allowing for a faster contraction of the envelope and thus $M_{\rm c}<M_{\rm env}$. The higher metallicity in pebbles then also allows for the formation of ice giants far out at late times in the disc, which was not that easy compared to the $Z=1.0\%$ case. 

Generally, the higher metallicity significantly broadens the parameter space that allows for forming gas giants. A high pebble accretion rate is required to form massive gas giants ($M_{\rm P} > M_{\rm Jup}$) that stay far outside in the disc ($r_{\rm f} > 10$~AU), which do not exist in the lower metallicity simulation. 

Observations also indicate that stars with higher metallicity host more giant planets in close orbits \citep{2004A&A...415.1153S, 2005ApJ...622.1102F}, which is confirmed by our results, where the formation of gas giants is possible in a wider range in the $r_{\rm 0}-t_{\rm 0}$ parameter space for high $Z$. If planet formation starts late (large $t_{\rm 0}$),  higher metallicity helps to form giant planets at late stages. Additionally, \citet{2014PNAS..11112655M} and \citet{2014Natur.509..593B} find that smaller planets ($R_{\rm P} < 4R_{\rm E}$) are slightly more common around stars with solar metallicity, which is reproduced by our lower metallicities simulations, where the formation of small planets in the $r_{\rm 0}-t_{\rm 0}$ parameter space is enhanced (top in Fig.~\ref{fig:Zenvelope}) compared to the high $Z$ simulations.

Even with changing metallicities in pebbles that allow for faster or slower growth of the core, gas giants are not able to form {\it \emph{in situ}}, because of their strong migration. On the other hand, the formation of ice giants results in much less migration through the disc, so these planets can form {\it \emph{in situ}}, but they have to form at late stages, because otherwise they would continue to grow to become gas giants.

\subsection{Amount of micrometre-sized dust in the disc}

The thermodynamic structure of the disc is determined through micrometre-sized dust grains. However, during the lifetime of the disc, the amount of micrometre-sized dust can change, for example because of grain growth. Larger grains do not contribute to the opacity, so that the smaller number of dust grains in the disc results in a colder disc, because cooling is increased \citep{2014arXiv1411.3255B}. This results in a smaller aspect ratio of the disc, which reduces the pebble scale height (eq.~\ref{eq:Hpebble}) and allows an earlier transition from 3D to 2D pebble accretion. Additionally, a smaller aspect ratio results in a larger $\tau_{\rm f}$, making the transition to 2D pebble accretion even smaller. Therefore planets in discs with a smaller amount of micrometre-sized dust grains can grow faster. However, a smaller aspect ratio of the disc reduces the pebble isolation mass (eq.~\ref{eq:Misolation}) resulting the outcome of the systems.

\begin{figure}
 \centering
 \includegraphics[scale=0.7]{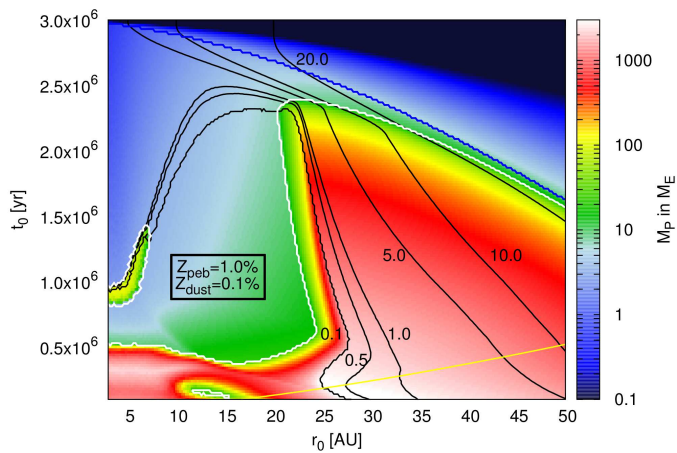}
 \caption{Final masses of planets ($M_{\rm P} = M_{\rm c}+M_{\rm env}$) as a function of initial radius $r_{\rm 0}$ and initial time $t_{\rm 0}$ in the disc. The plot features a metallicity of $Z=1.0\%$ in pebbles and an amount of $Z_{\rm dust} = 0.1\%$ in micrometer sized dust grains, which is a factor of $5$ smaller than in Fig.~\ref{fig:Z008kappa}. The different lines inside the plots have the same meaning as in Fig.~\ref{fig:Z008kappa}. A lower amount of dust grains results in a colder disc, which reduces the pebble isolation mass in the inner disc and therefore increases the parameter space that allows for the formation of ice giants compared to Fig.~\ref{fig:Z008kappa}. In the outer disc, formation of larger bodies is easier compared to Fig.~\ref{fig:Z008kappa}, because of the larger pebble size and smaller pebble scale height $H_{\rm peb}$ in the outer disc.
   \label{fig:Z010metal01}
   }
\end{figure}

We now discuss simulations with an amount of micrometre-sized dust of $Z_{\rm dust} =0.1\%$, which is five times smaller than in our nominal model (see Fig.~\ref{fig:Z008kappa}). All other parameters are the same as in our nominal model. Figure~\ref{fig:Z010metal01} shows a much larger parameter space in the inner parts of the disc that harbours ice giants than in the $Z_{\rm dust} = 0.5\%$ model. This is caused by the smaller aspect ratio in the inner disc, which resulted from the increased cooling \citep{2014arXiv1411.3255B}. This smaller aspect ratio results in a lower pebble isolation mass causing a longer contraction time of the gaseous envelope, which hinders the planets from reaching the runaway gas accretion stage.

Changing the disc structure also influences the migration of embedded planets. In Fig.~\ref{fig:Z008kappa} the region harbouring ice giants at $\sim 3$~AU is caused by a region of outward migration, which can contain planets of up to $\sim 8$ ${\rm M}_{\rm E}$ (Fig.~\ref{fig:Migcont}). However, in a disc with decreased $Z_{\rm dust}$ the region of outward migration can only contain planets of up to $\sim 3.5$ ${\rm M}_{\rm E}$. The size of the cores in this region is only $\sim 2.5$ ${\rm M}_{\rm E}$ owing to the low pebble isolation mass. The cores therefore need to contract a gaseous envelope of $\sim 2.5$ ${\rm M}_{\rm E}$ before they can undergo rapid gas accretion, which takes a very long time. This means that the planets can outgrow the region of outward migration, because they become too massive, while they still undergo a contraction phase of the envelope. They therefore migrate to the inner system as ice giants, making the formation of ice giants in the inner system ($r_{\rm f} < 1.0$~AU) possible, in contrast to the $Z_{\rm dust} = 0.5\%$ model. This is crucial because most exoplanets that have been observed have a few Earth masses and orbit very close to the central star \citep{2013ApJ...766...81F}. This area of parameter space can easily be populated in the case of low $Z_{\rm dust}$, which corresponds to cold protoplanetary discs.

In the outer parts of the disc, planet growth seems to be more efficient than in the $Z_{\rm dust} = 0.5\%$ model for two main reasons. The increased cooling results in a smaller aspect ratio in the outer parts of the disc, which increases the size of the pebbles (Eq.~\ref{eq:Stokesnumber}). Additionally, a larger pebble size reduces the scale height of the pebbles $H_{\rm peb}$. Both effects reduce the planetary mass needed to transition into the 2D pebble accretion branch (eq.~\ref{eq:2D3D}), resulting in a faster growth rate compared to $Z_{\rm dust} = 0.5\%$. This allows the efficient formation of gas giants in the outer disc, which then migrate into the inner disc. However, these planets will then only have a smaller planetary core than in the $Z_{\rm dust} = 0.5\%$ model, because the pebble isolation mass reduces as well with decreasing aspect ratio (eq.~\ref{eq:Misolation}).

A larger amount of micrometre-sized dust does not influence the outcome of our simulations significantly, because the changes in the disc structure are not as pronounced as for lower metallicities \citep{2014arXiv1411.3255B}.

\section{Planet formation via planetesimal accretion}
\label{sec:planetsimals}

In classic models of planet formation, planets grow via the accretion of planetesimals. The isolation mass for planetesimal accretion is different than when accreting pebbles. It is given by \citep{2002ApJ...581..666K, 2014prpl.conf..595R}
\begin{equation}
 M_{\rm iso,pla} = 0.16 \left(\frac{b}{10R_{\rm H}}\right)^{3/2} \left(\frac{\Sigma_{\rm pla}}{10}\right)^{3/2} \left( \frac{r}{1{\rm AU}}\right)^{1.5 (2-s_{\rm pla})} \left(\frac{M_\star}{M_\odot}\right)^{-0.5} {\rm M}_{\rm E} \ ,
\end{equation}
where $b$ is the orbital separation of the growing embryos, which we set to $10R_{\rm H}$. Here, $\Sigma_{\rm pla}$ is the surface density in planetesimals, and $s_{\rm pla}$  the negative gradient of the surface density in planetesimals. The accretion rate by planetesimal accretion is changed compared to pebble accretion. In \citet{2014arXiv1408.6087L}, the accretion rate for planetesimal is given by
\begin{equation}
\label{eq:planetesimals}
 \dot{M}_{\rm c,plan} =  \Psi \dot{M}_{\rm c,peb} = \Psi r_{\rm H} v_{\rm H} \Sigma_{\rm peb} \ ,
\end{equation}
where $\Psi$ is a reduction factor to the normal pebble accretion rate. It is given by
\begin{equation}
 \Psi = 3 \times 10^{-4} \left(\frac{r_{\rm p}}{10 {\rm AU}}\right)^{-1} \ .
\end{equation}
This follows directly from the assumption that the planetesimal velocity dispersion is equal to the Hill speed \citep{2009ApJ...707...79D,2010Icar..207..491D}, $v_{\rm H} = \Omega r_{\rm H}$, and gravitational focusing occurs from a radius $(r_{\rm c} r_{\rm H})^{1/2} r_{\rm H}$, which is smaller than the planetesimal scale height $H_{\rm pla} = v_{\rm H} / \Omega = r_{\rm H}$.

The general assumption of simulations with planetesimal accretion is that all solids in the disc are turned into planetesimals at the start of the simulation. We do the same here and point out that the surface density in planetesimals is $\Sigma_{\rm peb} = Z \Sigma_{\rm g}$, where we set $Z=8.0\%$, which is eight times higher than for the accretion with pebbles, and about eight times higher than the assumed metallicity of $1\%$ in the MMSN. With $Z=1.0\%$ planetesimal accretion is too slow, and no planets with $M_{\rm p}>1{\rm M}_{\rm E}$ are formed at all. Along with that, we keep the surface density of planetesimals constant through the evolution time of the disc, because planetesimals are safe from drifting through the disc by gas drag. This also means that we keep the initial gradients in the surface density of planetesimals that follow the initial gradients of the gas surface density $\Sigma_{\rm g}$. When the simulation starts at a later initial time $t_{\rm 0}$, we use a planetesimal surface density and gradient that follows $\Sigma_{\rm g}$ at that time $t_{\rm 0}$. When the planets reach their planetesimal isolation mass, they first have to undergo a contraction of the envelope before runaway gas accretion can start. However, since the time scale of forming cores with planetesimals is very long, we extend the disc's lifetime to $5$ Myr to extend the time for gas accretion. This is also because a disc lifetime of $3$ Myr did not produce any gas giants whatsoever.

\begin{figure}
 \centering
 \includegraphics[scale=0.69]{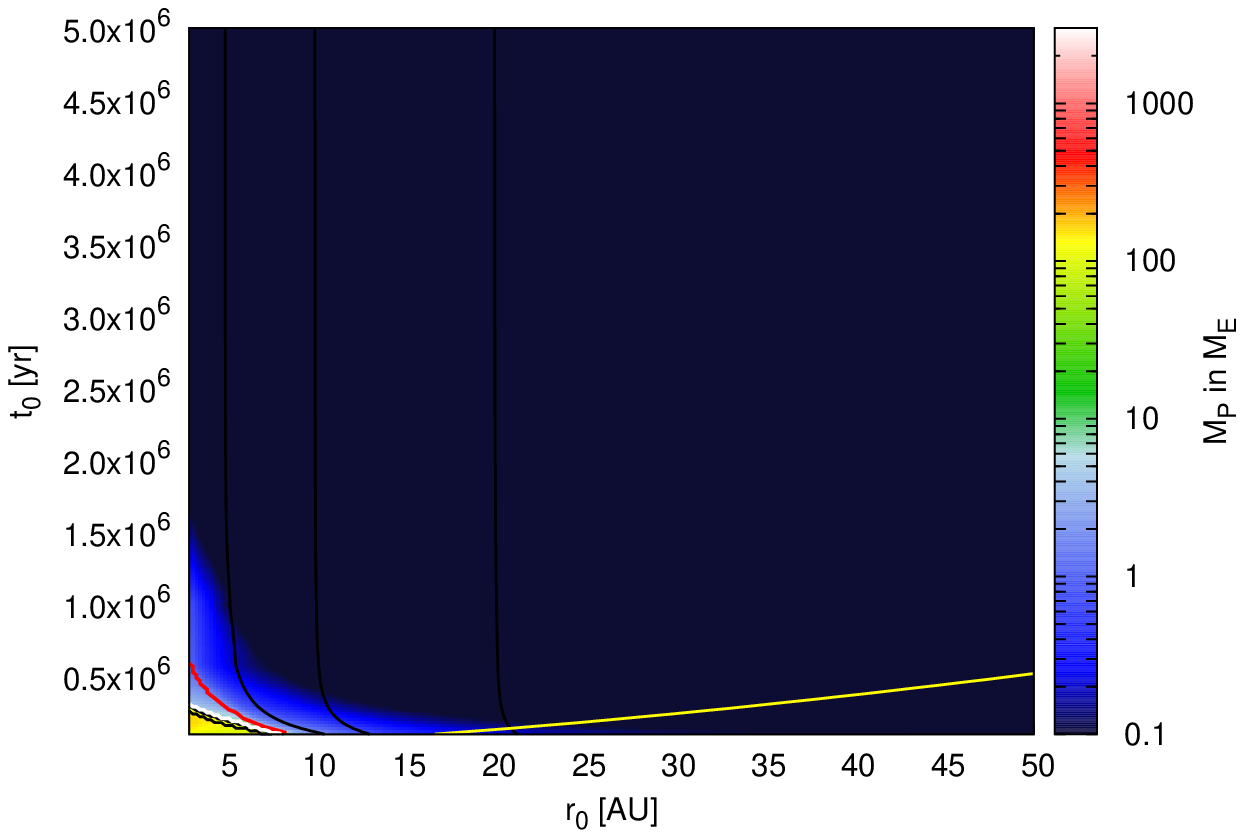}
 \includegraphics[scale=0.69]{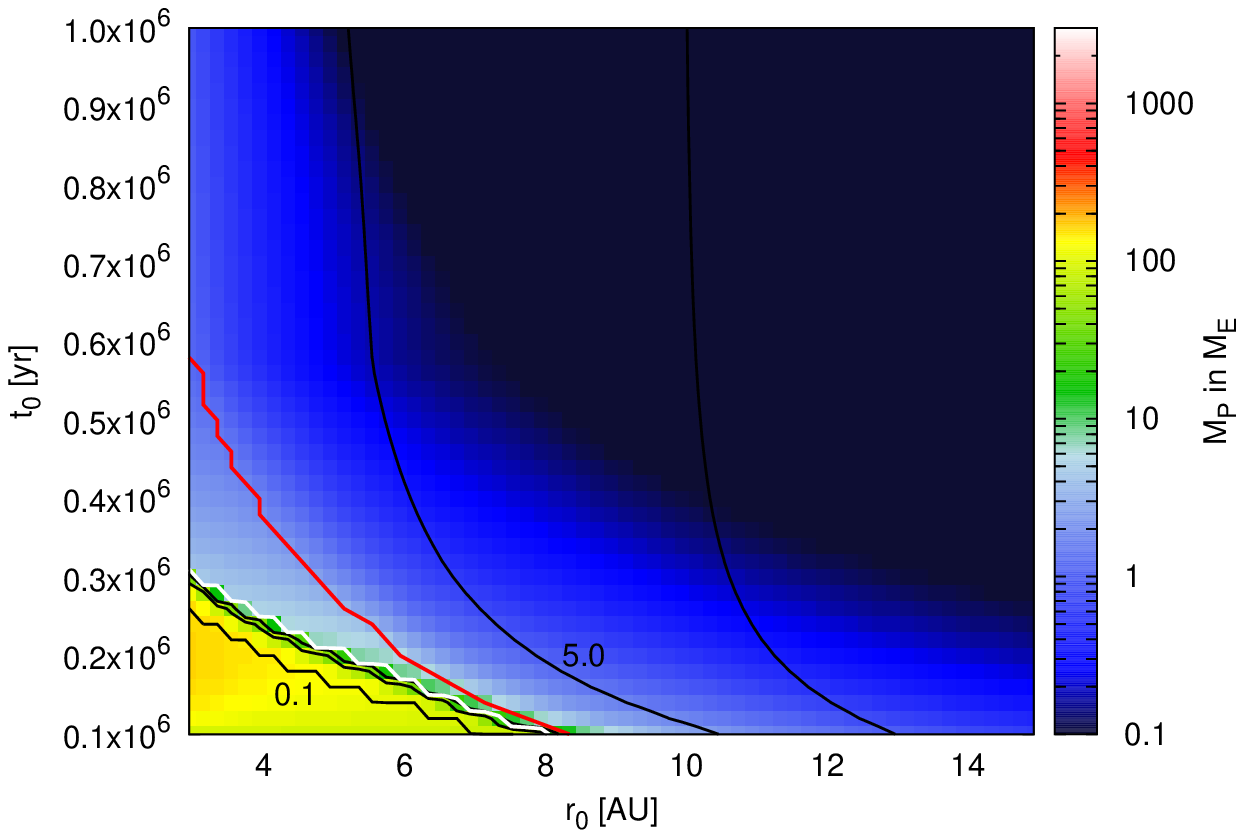}
 \caption{Final masses of planets forming by planetesimal accretion as a function of initial radius $r_{\rm 0}$ and initial time $t_{\rm 0}$ in the disc for $Z=8.0\%$ in planetesimals. The green line is the pebble production line, where we assume that a production of pebbles is needed to first form the planetesimals, so every point beneath the yellow line should not be taken into account. The red line now indicates the planetesimal isolation mass (eq.~\ref{eq:planetesimals}), so everything below it has reached planetesimal isolation mass. Everything above the white line has $M_{\rm c}<M_{\rm env}$. The top plot shows the full $r_0$-$t_0$ parameter space, while the bottom plot is zoom into the inner $r_0$-$t_0$ parameter space to enhance the interesting parts of the diagram.
   \label{fig:planetesimal}
   }
\end{figure}

In Fig.~\ref{fig:planetesimal} we display the final planetary masses for simulations where the planetary cores grow by the accretion of planetesimals with a metallicity of planetesimals of $Z=8.0\%$. The growth of planets, even with eight times higher metallicity is very slow, and only those planets that start to form early (small $t_{\rm 0}$) and in the inner regions of the disc (small $r_{\rm 0}$) reach planetesimal isolation mass. However, the planets that reach planetesimal isolation mass are very small and take a very long time to contract their envelope, which means they will migrate for a very long time in type-I migration. But the planets are indeed small enough, and in the inner regions of the disc these planets can get trapped in the zero-migration zones (Fig.~\ref{fig:Migcont},) allowing them to stay at a few AU from their host star before runaway gas accretion sets in. The planets then outgrow the region of outward migration and move towards the inner disc. In the end most of these planets are not massive enough to open up a gap in the disc, and they migrate inwards very fast. In fact, in this simulation none of the planets reach a mass that is comparable to Jupiter's. Additionally, these simulations fail to produce planets with $M_{\rm P}> 10{\rm M}_{\rm E}$ outside of $8$~AU.

With just the accretion of planetesimals it is very hard to form the cores of gas giants that stay out at $\sim 5$~AU, as does Jupiter in our own solar system. However, if multiple planetary embryos were present in the disc, these could collide and form bigger objects and eventually the cores of giant planets at these orbital distances \citep{2014arXiv1407.6011C}. With single embryos that accrete the planetesimals, effective growth to reach the stages of giant planets is not possible at all.

The reason we do not produce giant planets at a few AU in contrast to population synthesis models lies in their simpler disc model. The population synthesis models use a steeper gradient in planetesimal surface density, allowing for more planetesimals in the inner parts of the disc \citep{2008ApJ...673..487I}. A higher density of planetesimals then results in a faster growth rate of the cores, which also allows for a faster contraction of the gaseous envelope, giving the planet more time to reach the runaway gas accretion stage. Still, formation of gas giants in orbits beyond $5$~AU would not be possible, even under these generous conditions.

\section{Formation of the solar system}
\label{sec:solarsystem}

We now focus on the formation of the giant planets in our solar system, where we follow two different approaches. We first want to reproduce the planetary configuration at the start of the Nice model \citep{2005Natur.435..459T}, and in a second attempt we want to reproduce the giant planet configuration at the beginning of the Grand Tack scenario \citep{2011Natur.475..206W}, where we end our simulations just before the two gas giant planets start their outward migration in resonance. We stick here to our usual assumption of a metallicity of $1.0\%$ in pebbles that can be accreted onto the planet. 

When multiple planets form in the disc, the outermost pebble accreting planet will reduce the flux of pebbles seen by the inner planets by the amount that it accretes. However, this reduction of the pebble accretion stream is not significant when just considering four bodies \citep{2014arXiv1408.6087L}.

\subsection{Nice model}

The Nice model aims to explain the bombardment history of the inner solar system, roughly $1$ Gyr after the formation of the system \citep{2005Natur.435..459T}. In this model, the four giant planets in the system start in compact orbits after the gas disc dispersed, where Jupiter and Saturn are in a 3:2 resonance, and outside the ice giants is a belt of planetesimals containing roughly $30$ Earth masses. The system then becomes unstable, because of the constant scattering of planetesimals on the planets. This instability leads to an inward motion of Jupiter, while the other planets move outwards. In about $50\%$ of the simulations, Neptune and Uranus switch places \citep{2005Natur.435..459T}.

\begin{figure}
 \centering
 \includegraphics[scale=0.7]{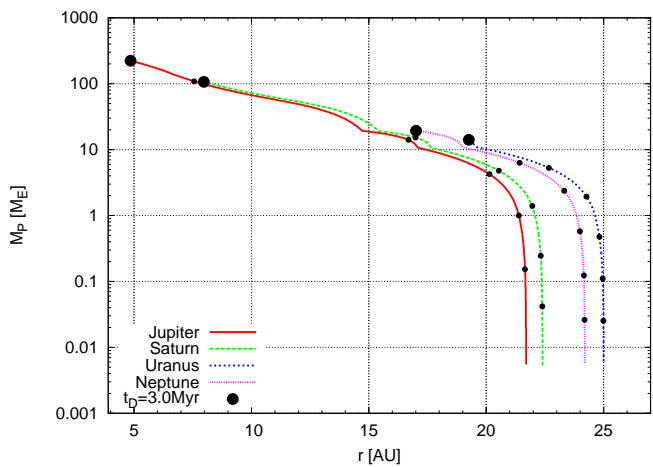}
 \caption{Evolution of the total planetary masses as a function of orbital distance for a proto-solar system with the giant planets ending up in a configuration similar to the initial conditions of the Nice model. Jupiter starts at $t_{\rm 0}=1.65$ Myr, Saturn at $t_{\rm 0}=1.77$ Myr, and Neptune and Uranus both at $t_{\rm 0}=1.81$ Myr. The big black circles label a total disc evolution time of $3.0$ Myr, while the small black dots indicate $2.0$, $2.2$, $2.4$, $2.6$, and $2.8$ Myr. We display Neptune inside of Uranus here, since they often switch places during the Nice model \citep{2005Natur.435..459T}. We used $Z_{\rm dust} = 0.5\%$ and a metallicity of pebbles of $Z=1.0\%$ as in our nominal disc model.
   \label{fig:Solarsystem}
   }
\end{figure}

We now aim to reproduce the masses and planetary orbits of the four giant planets in the solar system using our disc evolution and pebble accretion model. We keep the total disc lifetime at $3$~Myr, but also allow different formation times for the planetary seeds of the giant planets. The results are shown in Fig.~\ref{fig:Solarsystem}, which shows the final evolution of the system at $t_D=3$ My. The starting times of the planetary seeds differ between the planets. The final planetary masses and orbits needed for the starting configuration of the Nice model are reproduced quite well.

The formation of a gas giant is easy in the framework of the evolving disc and pebble accretion. However, the formation of the two gas giants of the solar system is more somewhat trickier, because in the final configuration, they differ by a factor of $3$ in mass, but are only $3$~AU apart from each other. The reproduction of this exact configuration is challenging because either Jupiter is slightly too small or Saturn is too big. This is a consequence of the interplay between pebble accretion, planet migration, gas envelope contraction, and runaway gas accretion. In the simulation shown here, our Jupiter analogue is only $\sim 230$ Earth masses. This difference compared to the real Jupiter can be caused by the slight uncertainties in our model, for example in the contraction of the gaseous envelope (see Appendix~\ref{ap:kappa}). Even though there is a time difference between the formation of Jupiter and Saturn, their orbits also start very close to each other, which might result in some minor interactions between the planets that is not taken into account here. Nevertheless, the reproduction of Jupiter's and Saturn's masses and orbital distances is remarkably good.

The formation of the ice giants that are in the mass range of Uranus and Neptune with a very low gas content is difficult because a larger core can attract gas more easily (eq.~\ref{eq:Mdotenv}), so the planet can very easily grow to become a gas giant. This limits the parameter space in $r_0$-$t_0$ that hosts ice giants with low gas content in the mass range of Uranus and Neptune at final orbital distances of $15$-$20$~AU (Fig.~\ref{fig:Z008kappa}). However, in a recent study, \citet{2015arXiv150603029I} have found that the formation of Uranus and Neptune by a series of giant impacts of planetary embryos of a few Earth masses outside of Saturns' orbit is possible. These planetary embryos can easily be formed by seeds growing with pebble accretion. When taking also the possibility of giant impacts into account, the formation of ice giants like in our solar system seems very likely, especially at late formation times. Full-grown ice giants and the planetary embryos can form in orbital distances of $20$~AU to $30$~AU at $2$ Myr (Fig.~\ref{fig:Envelope2Myr}). The region where ice giants can form in the $r_0$-$t_0$ parameter space is also indicated in Fig.~\ref{fig:categories}. It is a vast region in the parameter space.

We also checked that a change in $Z_{\rm dust}$ influences our results when reproducing the initial configurations of the Nice model, but found no such dependency. Reproducing the exact configuration of the solar system is very sensitive to the initial orbital distance and formation times of the planetary seeds, since those parameters determine the final planetary mass and orbital distance. The general outcome of having two gas giants, and outside of them, two ice giants can be reproduced very easily.

The orbital configuration of the solar system in the present day is different from the initial conditions of the Nice model. The ice giants are now much farther away from the Sun than in the initial configuration of the Nice model. In particular, Neptune is located $30$~AU from the Sun. Forming a Neptune-sized body with $r_{\rm f} = 30$~AU in our planet growth scheme is possible, but the planetary seed would then have to start deep within the Kuiper belt at $\sim 45$~AU and with an early formation time $t_0\approx 670$ kyr. When the planetary seed reaches $40$~AU, it has already grown to several Earth masses, which would have disrupted the Kuiper belt. This supports the Nice model concept of forming the giant planets in a close resonant configuration, because it leaves the Kuiper belt untouched.

\subsection{Grand Tack scenario}

The Grand Tack scenario \citep{2011Natur.475..206W} describes a scenario where Jupiter and Saturn migrate into the inner solar system (Jupiter down to $\sim 1.5$~AU) and then migrate outwards in resonance again. This outward migration of gap-opening planets in resonance was originally discovered by \citet{2001MNRAS.320L..55M}, but is applied to the solar system here. The appealing effect of the Grand Tack scenario is that the masses and orbital distances of the terrestrial planets, especially Earth and Mars, can easily be reproduced, as can the features of the asteroid belt \citep{2011Natur.475..206W}. Here we now want to reproduce the masses and orbital distance of Jupiter and Saturn before both planets begin their outward migration in resonance.

Because we only model single planets in the disc, we try to arrive at a configuration that would allow for the outward motion of the giant planets in the gas disc in resonance. Additionally, because the outward migration happens in a gas disc, the planets will still accrete gas during this outward motion. We therefore aim to have a Jupiter analogue with $\sim 200$ ${\rm M}_{\rm E}$ and a Saturn analogue with $\sim 70$ ${\rm M}_{\rm E}$. This is much smaller than their current mass, but during their outward migration in the gas disc, Jupiter and Saturn can still accrete gas and reach their final mass. We also aim for stranding the planets roughly in a mutual resonance, where we put Jupiter at $\sim 2$~AU. We try to put Saturn's orbit somewhere between the 2:1 and 3:2 resonance with Jupiter, because both resonances allow outward migration in discs with small $H/r$ \citep{2014ApJ...795L..11P}, as we have here. We also assume that when the disc reaches $3$ Myr of lifetime, the disc will still live long enough to allow the giant planets to migrate outwards and accrete the rest of their mass.

\begin{figure}
 \centering
 \includegraphics[scale=0.69]{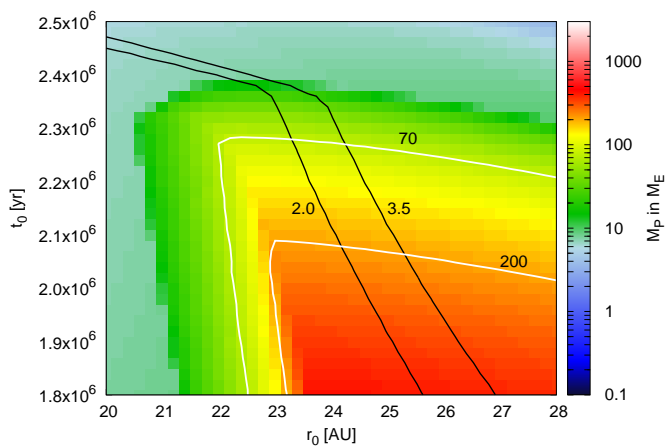}
 \caption{Configuration of our assumptions of the Grand Tack scenario in a disc with $Z_{\rm dust} = 0.1\%$. The black lines indicate the orbital distances of $2.0$ and $3.5$~AU, which correspond to the $2:1$ mean motion resonance when the outward migration in the Grand Tack scenario can start. The white lines correspond to the final planetary masses of $200$ ${\rm M}_{\rm E}$ for Jupiter and $70$ ${\rm M}_{\rm E}$ for Saturn. Jupiter's $r_0$ and $t_0$ are therefore constrained by the crossing between the white $200$ ${\rm M}_{\rm E}$ and the black $2$~AU lines, while Saturn is constrained by the white $70$ ${\rm M}_{\rm E}$ line and the black $3.5$~AU line. Jupiter forms earlier (smaller $t_0$) and closer to the Sun (smaller $r_0$) than Saturn, showing that the conditions of the Grand Tack scenario can be met.
   \label{fig:GT}
   }
\end{figure}

Figure~\ref{fig:GT} shows the $r_0$-$t_0$ parameter space of interest for the formation of Jupiter and Saturn with the above-mentioned characteristics in a disc with $Z_{\rm dust} = 0.1\%$. The initial $r_0$ of Jupiter and Saturn is quite close to each other, but they are separated in $t_0$ by nearly $200$ kyr. During this time, Jupiter's core grows to a few tenths of an Earth mass and migrates only a minimum distance through the disc. However, before the mass of Jupiter becomes massive enough to disturb the orbit of Saturn's seeds, it migrate a few AU from its initial position.

In the Grand Tack scenario proposed by \citet{2011Natur.475..206W} Saturn migrates much faster than Jupiter and catches it. This is explained by type-III migration. However, our Jupiter and Saturn seeds are both placed in a region of parameter space that is subject to type-III migration (see appendix~\ref{ap:typeIII}). This indicates that Jupiter must stop its inward type-III migration and transition into type-II migration, so that Saturn catches up. However, this process is not investigated in full. Likewise, the outward migration of Jupiter and Saturn in resonance is thought to end when the aspect ratio of the disc increases \citep{2001MNRAS.320L..55M}. However discs with low numbers of dust grains at the late evolutionary stages have low aspect ratios to greater distances than Saturns current orbit \citep{2014arXiv1411.3255B}.

However, the formation of Jupiter and Saturn for the configuration of the Grand Tack scenario is only possible in discs with a small number of dust particles. We did not find not find any allowed parameter-combination in $r_{\rm 0}-t_{\rm 0}$ space for discs with large amounts of micrometre-sized dust ($Z_{\rm dust} > 0.3\%$) that would allow for the starting configuration of the Grand Tack scenario within our assumptions above. Given the uncertainties in our model (e.g. gas envelope contraction), more investigations are certainly needed to reproduce the initial configuration of the giant planets in the Grand Tack scenario.

\section{Discussion}
\label{sec:discussion}

\subsection{Evolution and lifetime of the disc}

The evolution of the disc is coupled to the accretion rate, where we follow the observations of \citet{1998ApJ...495..385H}. We make the assumption that when the accretion rate is very low, $\dot{M} < 2 \times 10^{-9} M_\odot$/yr, photoevaporation clears the disc very rapidly \citep{2013arXiv1311.1819A}, and we end our simulation at that stage. This results in a lifetime of $3$ Myr. If the disc evolution continues to $\dot{M} = 1 \times 10^{-9} M_\odot$/yr, the lifetime would be $5$ Myr. This additional lifetime of $2$ Myr influences the results because in the late stages of the disc, the disc mass is low so the pebble accretion rate onto the planet is low as well. The planets therefore no longer grow efficiently, but at the same time the planets can migrate significantly in the disc, because migration is mostly inwards at that stage of the disc's lifetime. Combined with the long time scale of $5$ Myr, planets can only stay outside of $0.1$~AU if they form late (after $\sim 2.5$ Myr) and far outside in the disc. In the study by \citet{2009AIPC.1158....3M} only $\sim 30\%$ of all stars are found to still harbour protoplanetary discs after $2.5$ Myr, so our disc lifetime of $3$ Myr might be even on the longer side. Different disc lifetimes are discussed in more detail in Appendix~\ref{ap:decay}.

The evolution and formation of planets inside the disc depends crucially on the disc structure and its evolution. If the disc can sustain higher $\dot{M}$ rates for a longer time, planet formation is significantly different, because the accretion rates will be higher and the disc might entertain regions of outward migration for higher planetary masses for a longer time, allowing more planets to stay farther out in the disc. Improved observational constraints on the lifetimes of protoplanetary discs would therefore be very useful for understanding planet formation better.

\subsection{Inner disc}

We stop the inward migration when the planet reaches $r_{\rm f}<0.1$~AU, which is the inner edge of the disc. We then stop the whole evolution of the planet at that point, because tidal effects so close to the star can become important, which we do not consider here.

In additional, planetary growth that close to the central star might work differently; for example, the pebbles that can be accreted from planets that are at $r_{\rm P} < r_{\rm ice}$ are significantly smaller than for $r_{\rm P} > r_{\rm ice}$. Silicate particles will bounce off each other instead of growing \citep{2010A&A...513A..57Z}, keeping the particles small. We also stop the growth of the planet, because gas accretion in the inner hot parts of the disc might work completely differently from gas accretion in the outer colder parts of the disc, because the surrounding gas temperature is higher, making cooling less efficient, thereby prolonging the contraction phase of the envelope. The final mass of the planets inside the parameter space $r_{\rm 0}$ and $t_{\rm 0}$, which allows for $r_{\rm f}<0.1$, is therefore not $100\%$ accurate, but orbital evolution is taken into account self-consistently until the planets reach $0.1$~AU.

An additional challenge emerges when looking at results from N-body simulations, where the inner edge of the disc acts as a halting zone for inward migration. In \citet{2014arXiv1407.6011C} the inner edge acts as small zone of outward migration that stops planets. These planets can then act as a buffer zone for other embryos migrating into the inner disc and allow pile-ups in resonances. The planets trapped in resonance can grow more and become bigger planets farther out in the disc. Also, in our simulations, only gas giants migrate to sub-AU orbits, while smaller ice giants remain around their birth locations.

\subsection{Convergence zones}

In the inner regions of the disc, where $H/r$ drops with radius, zones of outward migration exist. In the early stages of the disc evolution, these zones of outward migration are several AU wide and can harbour planets from $5$ to $40$ Earth masses. However, as the disc accretes in time, the regions of outward migration shrink and can only contain planets with lower mass (see Fig.~\ref{fig:Migcont}). At an age of $\sim 1$ Myr the regions of outward migration exist only in the inner few AU and for planets with $M_{\rm P} < 10$ ${\rm M}_{\rm E}$. 

This implies that planets that form early in the disc will outgrow the regions of outward migration owing to gas accretion and migrate to the inner disc. In the early stages of the disc evolution, the region of outward migration allows the planets to grow to several tens of Earth masses before they migrate to the inner disc, but it does not prevent the planets from migrating to the inner disc. It just delays their inward motion.

In the late stages of the disc evolution, in combination with the slow gas accretion due to the long contraction time of the envelope, however, the region of outward migration is crucial. Late-forming planets ($t_{\rm 0}>2$ Myr) in the inner disc have only a very small core, because the pebble isolation mass is low ($\sim 3 {\rm M}_{\rm E}$). Therefore the contraction of the envelope takes a long time (eq.~\ref{eq:Mdotenv}), which hinders the planet from outgrowing the region of outward migration and lets it stay at a few AU from the central star. These planets in the inner disc then classify as ice giants, because they are just a few Earth masses, their core mass is higher than their envelope mass, and they form at $r_{\rm P}>r_{\rm ice}$.

Additionally, the {\it \emph{in situ}} formation of giant planets is not possible when we take migration into account. Even planets that have a final orbital distance of a few AU have most likely not formed there, but migrated inwards from even farther out (see Fig.~\ref{fig:Z008kappa}). Smaller planets, on the other hand, can form {\it \emph{in situ}}, but only in orbits beyond $1$~AU.

In \citet{2014arXiv1408.6094L} the planets form in an inside-out fashion, where planets closer to the star form before planets farther out, when the seed cores from around the same time in the disc. However, inside ten AU, this simple mechanism breaks down in our model. Time-evolving discs including convergence zones for migrating planets do not allow such a simple picture any more (see Figs.~1 and~\ref{fig:Z008kappa}).

\subsection{Formation of the solar system}

The formation of the exact configuration of our own solar system is relatively straightforward. The formation of the cores of the giant planets can easily be explained with pebble accretion \citep{2014arXiv1408.6087L}, however reproducing the exact planetary masses and orbital distances requires some fine-tuning of the model. The two main problems are (i) that there are unknown opacities in the planetary atmosphere when the gaseous envelope contracts, which changes the time scale of gas envelope contraction; and (ii) that we were just considering single planets in the disc, meaning that we do not consider giant impacts for the formation of the cores. A study by \citet{2015arXiv150603029I} explains the formation of Uranus and Neptune through the accretion of planetary embryos of a few Earth masses, which can be formed easily by pebble accretion. Taking this growth mechanism into account makes the formation of the ice giants of our solar system even easier.

Reproducing the initial conditions for the Grand Tack scenario \citep{2011Natur.475..206W} is more difficult because only discs with a low amount of dust particles (low $Z_{\rm dust}$) allow reproduction of initial conditions that can lead to the Grand Tack scenario. The problem faced is the necessarily big mass difference between Jupiter and Saturn even though they are quite close in their orbital separation at the beginning of their combined outward migration. During their outward migration, Jupiter and Saturn can still accrete gas, which would allow them to grow, but making it very hard to constrain their initial mass at the start of their outward migration. This scenario should be investigated in more detail with simulations that feature multiple cores.

The formation of the orbital configuration needed for the late instability in the Nice model \citep{2005Natur.435..459T} can be reproduced quite well within our simulations, where the amount of dust particles for the disc structure does not play an important role. The reproduction of the Nice model configuration is easier because larger orbital distances between the giant planets are needed than in the Grand Tack scenario, which allows the mass ratio between the giant planets to be reproduced. This is much harder in the Grand Tack scenario (Fig.~\ref{fig:GT}). The orbital distances of the giant planets are matched for our Nice model configuration, while only the mass of Jupiter is slightly off (see Fig.~\ref{fig:Solarsystem}). Our model is therefore very successful in reproducing the configuration of the giant planets in our solar system.

\subsection{Formation of Jupiter}

The atmosphere of Jupiter is enriched in noble gases compared to the solar values \citep{1999Natur.402..269O}, which led to the conclusion that the planetesimals accreted by Jupiter formed in the cold regions of the disc \citep{1999Natur.402..269O}. In this scenario the formation of Jupiter's core starts at $5$~AU, and planetesimals from the cold region of the solar system are accreted in order to provide the enrichments in noble gases. \citet{2006MNRAS.367L..47G} proposed that the accretion of Jupiter's noble gases occurs at a late stage of the disc evolution, probably during the times when photoevaporation of the disc already has started because the temperature is low and noble gases can condense out.

However, our Jupiter analogue (Fig.~\ref{fig:Solarsystem}) starts to form at $20$~AU and reaches pebble isolation mass when it has migrated down to $\sim 17$~AU. Since the formation starts at $t_{\rm 0}=2$ Myr, the surrounding disc temperature in that region is already very cold ($T\approx 30$K), which allows for the condensation of ices that engulf noble gases and which then would form the core of Jupiter. Moreover, the accretion of gas onto the core also happens in the cold regions of the disc, which could further explain the enrichment of noble gases in Jupiter's atmosphere. Our model implies a late formation of Jupiter at a large distance and therefore would naturally explain the enrichment of noble gases in Jupiter's atmosphere.

Our model therefore also predicts a noble gas enrichment in Saturns atmosphere, because Saturn forms in the cold parts of the disc. Unfortunately, the noble gas enrichment of Saturn's atmosphere is unknown at the moment.

\subsection{Formation of hot gas giants}

The formation of hot gas giants close to the central star can happen through two formation channels. Either by inward migration \citep{1996Natur.380..606L, 1997Icar..126..261W, 2003ApJ...588..494M} or by scattering events \citep{1996Sci...274..954R, 2002Icar..156..570M}.
Inward migration of gas giants through type-II or type-III is reasonable possibility in our formation scenario, because the migration rates are very strong (Fig.~\ref{fig:Z008kappa}). However, when comparing the simulations with $Z=1.0\%$ and $Z=1.5\%$ in pebbles, it seems that the amount of parameter space needed to form gas giants with $r_{\rm f}<0.1$~AU is slightly smaller in the $Z=1.5\%$ case. This effect is caused by the faster growth rate of planets, making the planets in the $Z=1.5\%$ disc more massive than in the $Z=1.0\%$ case, which then results in a reduced type-II migration rate because the planets become so big that they slow down their inward migration (eq.~\ref{eq:typeII}). However, this difference in $r_0$-$t_0$ parameter space might be a bit deluded by type-III migration, which acts on more planets in the high $Z$ case (see appendix~\ref{ap:typeIII}), which in turn might lead to more planets reaching $r_{\rm f}<0.1$~AU in the high $Z$ case.

Scattering events are also used to explain the distribution of hot gas giants, especially for hot gas giants with high eccentricity \citep{2002Icar..156..570M, 2014prpl.conf..787D}. These scattering events can be caused by other giant planets far out in the system. This clearly favours the high $Z$ case, because the formation of giant planets at large orbits (up to $30$~AU) is possible easily until the disc has reached an age of $\sim 2.0$ Myr (Fig.~\ref{fig:Zenvelope}), while for the lower $Z$ cases it is hard to form giant planets even up to $\sim 15$~AU (Fig.~\ref{fig:Z008kappa}). The larger formation probability of gas giants at greater distances makes the scattering hypothesis more probable, because there can simply be more giant planets available that can be scattered to the inner system to form a hot gas giant.

The combination of those two possibilities (disc migration and scattering) supports observations that hot gas giants are found around host stars with higher metallicity \citep{2004A&A...415.1153S, 2005ApJ...622.1102F}. Many hot gas giants have a large content of heavy elements \citep{2013Icar..226.1625M}, which can be explained by scattering events of planets that formed far out in the disc, because the core mass in the outer disc is higher owing to the higher pebble isolation mass.

\subsection{Planetesimal accretion}

The accretion of planetesimals can be slower than the lifetime of the disc \citep{1996Icar..124...62P, 2004AJ....128.1348R, 2010AJ....139.1297L}, even when considering an increased amount of solids inside the disc. We confirm these results (Fig.~\ref{fig:planetesimal}) and show that pebble accretion is much more efficient, even at the lower metallicity of pebbles than for planetesimals (Fig.~\ref{fig:Z008kappa}). In our nominal disc that has $0.5\%$ of metallicity in micrometre-sized dust and $1.0\%$ of metallicity in pebbles to form the planets, we are able to form a solar system analogue, which is not possible with planetesimals.

Planetesimal accretion with a metallicity of $8.0\%$ in planetesimals is only able to explain the formation of gas giants in the inner regions of the disc ($r_{\rm f}<1$~AU), if along with the high metallicity, the lifetime is prolonged to at least $4$ Myr. This is another limitation, because at that age photoevaporation might clear the disc efficiently. Observations by \citet{2009AIPC.1158....3M} also indicate that only $\sim 30\%$ of all stars still harbour protoplanetary discs after an age of $2.5$ Myr. In our simulations with planetesimal accretion, we are not able to form a Jupiter planet at $5$~AU, in contrast to \citet{1996Icar..124...62P}, because the isolation mass is lower there. This in turn leads to a longer contraction time for the envelope to prevent the accretion of gas, because our protoplanetary disc models are much colder than the MMSN. Additionally, we included the migration of planets, which leads to inward migration of planets, so that it is not possible to form a Jupiter analogue at $5$~AU as in \citet{1996Icar..124...62P}.

In addition, planetesimal accretion is not able to form the cores of the ice giants in the outer system. Even with an enhancement of the planetesimal density by a factor eight, planetesimal accretion alone can therefore not explain the formation of the solar system.

\subsection{Ice planets}

Our simulations predict a new class of planets, the so-called ice planets. They have a more massive core than envelope, but only reach a few Earth masses, so they do not qualify as ice giants. We also call them ice planets because they form outside the snow line ($r_{\rm p} > r_{\rm ice}$), meaning they mostly consist of ices and some gas, so they should have a lower density than rocky planets. These ice planets can be formed in the late stages of the disc evolution ($t_{\rm 0}>2$ Myr) in the outer regions of the disc. In these parts of the disc, the planets do not reach pebble isolation mass, because there is not enough material available at the late stages of the disc evolution, hindering growth. If formation of planets takes place generally at later stages in the disc evolution, and if the disc can still provide pebbles at these late stages, ice planets should be very common and be detectable by observations at large distances around their host stars. These type of planets are not predicted, however, and are not seen by population synthesis studies \citep{2014A&A...567A.121D}, because the formation with planetesimals takes too long, even for these low masses (Fig.~\ref{fig:planetesimal}).

\subsection{Population synthesis}
\label{subsec:popsynth}

Population synthesis models attempt to reproduce the observed distributions of exoplanets by combining planetesimal accretion, gas accretion, and planet migration in protoplanetary discs. The first investigations of this nature started nearly a decade ago \citep{2004ApJ...604..388I, 2004A&A...417L..25A}. However, to make these models work, some fundamental assumptions are made.

{\it Firstly}, because the accretion of planetesimals is very inefficient \citep{2010AJ....139.1297L}, a much higher total amount of solids than the standard solar value has to be assumed in order to make the population synthesis simulations work. In \citet{2004ApJ...604..388I} the total amount of planetesimals has to be at least five times the solar value in order to produce planets effectively.

{\it Secondly}, the migration rates of protoplanets inside the discs are found to be too high. \citet{2008ApJ...673..487I} find that type-I-migration needs to be reduced by factors of $10-100$ to reproduce the observed distributions of exoplanets. However, by taking outward migration due to the entropy driven corotation torque \citep{2006A&A...459L..17P} into account, the reduction of type-I migration can be greatly decreased so that it might not be needed any more \citep{2013A&A...558A.109A, 2014A&A...567A.121D}. However, in these simulations $\sim 60\%$ of the formed planets still move inside $0.1$~AU \citep{2014A&A...567A.121D}, showing that migration is still a problem for planet population synthesis models \citep{2014prpl.conf..691B}.

{\it Thirdly}, the evolution of the disc is either considered only via a simple power law \citep{2004ApJ...604..388I, 2008ApJ...673..487I} or follows a viscous disc evolution, where the gradients in surface density and temperature do not change in value but only move inwards as the disc accretes \citep{2014A&A...567A.121D}. Keeping the gradient in surface density and temperature at the same value has important consequences for the torque acting on embedded planets in the disc, because these gradients determine the migration rates and direction of planets (see section~\ref{subsec:migration}). However, recent simulations of protoplanetary discs taking heating and cooling effects into account have shown that these gradients change as the disc evolves in time \citep{2014arXiv1411.3255B, 2015arXiv150303352B}.

{\it Fourthly}, the initial gas mass, metallicity, and lifetime of the disc and the initial starting position of the planetary seeds are randomized in population synthesis studies \citep{2014prpl.conf..691B}. This can allow for the formation of massive planets ($M_{\rm P} > 100 {\rm M}_{\rm E}$) in the outer regions of the disc ($r_{\rm P} \approx 10$~AU), where the random selection of a very high disc mass allowed the formation of massive planets even in the outer disc. Disc column densities of more than ten times the MMSN are not found in observations of real protoplanetary discs \citep{2011ARA&A..49...67W}. However, even for such high disc masses, population synthesis models have problems forming planets at $r_{\rm P} > 10$~AU.

After applying these fundamental assumptions to their models, population synthesis models can reproduce the observed distribution of exoplanets to some extent. However, they still face two challenges: i) a huge amount of the initial seed planets migrate to the inner edge of the disc and ii) they fail to reproduce ice giant planets in the outer parts of the disc ($r_{\rm f}>8$~AU) on a regular basis. These type of planets are only produced if the conditions for the surface density of solids and disc dispersal time are just right by coincidence.

In the pebble accretion scenario we do not make any simplifications or reductions regarding the migration speed of planets, nor do we increase the amount of solids in our disc to unreasonably high amounts. Our planets can form with the nominal solar metallicity, where we use $0.5\%$ of solids in micrometre-sized dust grains and $1.0\%$ of solids in the form of pebbles, which makes a total of $1.5\%$ in solids.

The amount of planets that reach the inner edge of the disc is greatly reduced when the planets start to grow late in the disc. Considering that only a few percent of the stars harbour hot gas giants, a late formation scenario of planetary seeds ($t_{\rm 0} \approx 2$ Myr for disc lifetime of $3$ Myr) with pebble accretion naturally explains this observation, because the available time until disc dispersion is reduced, making the formation of hot gas giants more difficult. The late formation of planets also results in the formation of ice giants in the inner disc (between $1$ and $5$~AU). Those are amongst the most detected planets in our galaxy \citep{2013ApJ...766...81F}.

Pebble accretion is very efficient in the outer disc in the early stages (Fig.~\ref{fig:Z008kappa}), but becomes less efficient as time evolves. However, pebble accretion is much more efficient than planetesimal accretion, which allows the formation of ice planets in the outer disc.

In contrast to population synthesis models, where the outcome (which type of planet) is determined by the initial conditions such as the mass and lifetime of the protoplanetary disc and the surface density of planetesimals, our pebble accretion approach shows that the outcome of planet formation can be very different depending on the different initial times at which the planetary seed is placed in the disc.

We have also tested whether planets in our simulations are subject to type-III migration \citep{2003ApJ...588..494M} or are influenced by dynamical corotation torques (see Appendix~\ref{ap:typeIII}). In our simulations, we find that these effects only play a role in the early stages of the disc evolution, when the disc is still massive. At the later stages of the disc evolution, these effects no longer play a significant role because the disc is not that massive any more. 

The effect of type-III migration and of the dynamical corotation torques are crucial for population synthesis models, where a higher disc mass is needed to form planets. However, population synthesis models do not take type-III migration into account, which makes evolution of the semi-major axes in population synthesis models very doubtful, because type-III migration and dynamical corotation torques can lead to a faster inward migration. This can result in even more planets that are lost to the inner disc in population synthesis models.

\section{Summary}
\label{sec:summary}

In this paper we have combined previous results and theories of planet migration \citep{1997Icar..126..261W, 1986ApJ...309..846L, 2006A&A...459L..17P} with pebble accretion \citep{2010A&A...520A..43O, 2012A&A...546A..18M, 2012A&A...544A..32L, 2014arXiv1408.6094L, 2014arXiv1408.6087L} and gas accretion \citep{2010MNRAS.405.1227M, 2014ApJ...786...21P} in evolving protoplanetary discs \citep{2014arXiv1411.3255B} around Sun-like stars. We aimed to study where and when planets in a disc have to form in order to achieve a certain planetary mass and orbital distance to the star, which we used to define some distinct planetary types.

The initial planetary seeds grow via pebble accretion until they reach their pebble isolation mass \citep{2014arXiv1408.6087L}, which is when gas accretion can start. This initial growth stage depends on the metallicity of pebbles $Z$ in the disc, i.e. the number of pebbles available to be accreted, and on the location in the disc, because the pebble isolation mass depends on the aspect ratio of the disc (eq.~\ref{eq:Misolation}). The latter changes in time as the disc evolves \citep{2014arXiv1411.3255B}. As the aspect ratio of the disc decreases in time, the final mass of the cores become lower for planets that start their evolution in the disc at a later time $t_{\rm 0}$. Likewise, the mass of the core increases for planets that start to form at greater orbital distances $r_{\rm 0}$, because of the flaring structure of the disc, where $H/r$ increases with increasing $r$ and thus increases the pebble isolation mass.

After the planet reaches pebble isolation mass, it can start to accrete gas. But before runaway gas accretion can start, the gaseous envelope has to contract to reach $M_{\rm c} < M_{\rm env}$. This contraction time depends critically on the opacity inside the planetary atmosphere, which is not very well constrained, therefore future work should aim to study gas accretion and envelope contraction of planets with gaseous atmospheres in a much better way. After the envelope contracted, runaway gas accretion can start and the planets can become gas giants.

During the growth process of the planets, they migrate through the disc. While they accrete pebbles, the planets are small and do not significantly disturb the surroundings of the disc, meaning they migrate in type-I migration. This migration is inwards in the outer regions of the disc and can be outwards in the inner regions of the disc. However, these regions of outward migration change as the disc evolves and accretes onto the star. When the planet starts to accrete gas, it can also outgrow the regions of outward migration, and the planet moves fast inwards towards the star. But, when the planet becomes massive enough, it will start to open a gap in the disc and migrate in the slow type-II migration, which can keep it from reaching the inner edge of the disc.

\begin{figure}
 \centering
 \includegraphics[scale=0.69]{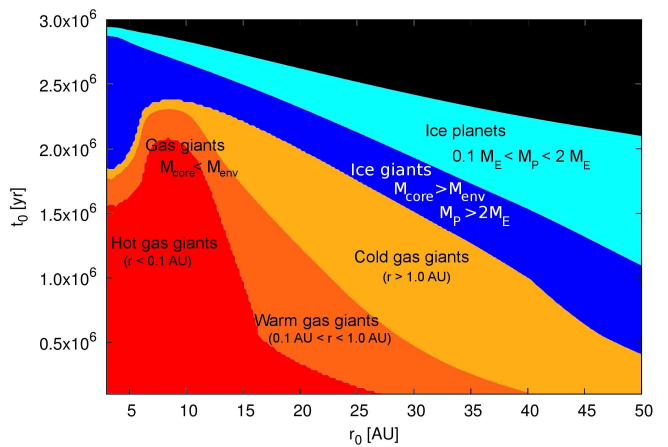}
 \caption{Categories of planets as a function of $r_{\rm 0}$ and $t_{\rm 0}$. The different colours indicate the different types of planets, which are separated by planetary mass and their core-to-envelope mass ratio. The hot, warm, and cold gas giants are separated by their semi-major axis (see Table~\ref{tab:classes}). The black region contains small objects with $M<0.1 {\rm M}_{\rm E}$.
   \label{fig:categories}
   }
\end{figure}

We can roughly categorise the planets by their their final mass $M_{\rm P}$ and orbital distance $r_{\rm f}$ in the disc. The formation regions of these classes are displayed in Fig.~\ref{fig:categories}.

{\it Hot gas giants} form early in the inner regions of the disc (small $t_{\rm 0}$ and small $r_{\rm 0}$), grow quickly to the pebble isolation mass, and can then be trapped in a region of outward migration, where their envelope contracts until they can accrete gas in a runaway fashion. At that point they have outgrown the region of outward migration and migrate into the inner disc, where they get stranded close to the central star.

{\it Warm} and {\it cold gas giants} have a much wider parameter space in $r_{\rm 0}-t_{\rm 0}$ to form than hot gas giants. They can form early far out in the disc, while they can still form later in the disc (up to $t_{\rm 0} \approx 2.2$ Myr) at moderate distances ($r_{\rm 0} \approx 15$~AU). They undergo the same kind of growth mechanisms as hot gas giants, but because they form initially at a greater orbital distance or a later time, they need to migrate more to end up in the inner disc as a hot gas giant. Their larger initial semi-major axis or late formation time keeps them far out in the disc.

{\it Ice giants} are planets located in the dark blue region in Fig.~\ref{fig:categories}. They have $M_{\rm c} > M_{\rm env}$ and a total mass that is higher than $2$ Earth masses. They can form at all orbital distances, depending on their formation time. The formation in the outer disc allows them to have a larger core, because of the large pebble isolation mass, and the late formation time hinders an efficient contraction of the envelope, so they do not reach runaway gas accretion. In the inner parts of the disc, the pebble isolation mass is lower (low $H/r$), which prolongs the contraction of the envelope, resulting in a lot of planets with $M_{\rm c} > M_{\rm env}$, which are ice giants by our definition.

{\it Ice planets} are small planets that have formed in the cold parts of the disc $r_{\rm f} > r_{\rm ice}$. These planets are very common in the outer regions of the disc, if the growth of planets starts in the late stages of the disc evolution ($t_{\rm 0}>2$ Myr). These planets did not reach pebble isolation mass and have therefore only a very minimal gaseous envelope, which should allow them to have a density similar to ice giants.

In contrast to previous simulations of the formation of giant planets by planetesimal accretion, we do not make any simplifications regarding planet migration, nor do we have to assume an unreasonably high amount of solids in the protoplanetary disc to form planets \citep{1996Icar..124...62P}. Additionally, our protoplanetary disc evolves in time and changes not only its total mass, but also its temperature and density profiles as the accretion rate decreases. Therefore, the pebble accretion scenario suggests that different types of planets emerge as a result of different formation time and location in the disc.

We have shown here that pebble accretion can overcome many of the challenges in the formation of ice and gas giants in evolving protoplanetary discs. Gas giants in our model do not form {\it \emph{in situ}}, but migrate over several AU during their formation process, which requires a formation of the planetary seeds far out in the disc. In contrast to that, ice planets and ice giants are more likely to form {\it \emph{in situ}} compared to giant planets. In fact, ice planets and ice giants only form in the late stages of the disc evolution or in the outer parts of the disc. A late formation time thus gives rise to a wide variety of planetary types, akin to those found in our solar system, as well as in extrasolar planetary systems.

\begin{acknowledgements}

B.B.,\,M.L.,\,and A.J.\,thank the Knut and Alice Wallenberg Foundation for their financial support. B.B.\, also thanks the Royal Physiographic Society for their financial support. A.J.\,was also supported by the Swedish Research Council (grant 2010-3710) and the European Research Council (ERC Starting Grant 278675-PEBBLE2PLANET). We thank the referee John Chambers for his comments that helped to improve the manuscript. We thank F. Masset for his helpful discussions regarding the heating torque.

\end{acknowledgements}

\appendix
\section{Dispersal time of the disc}
\label{ap:decay}

Observations have shown that the lifetime of protoplanetary discs is a around a few Myr \citep{1998ApJ...495..385H}. However, the fraction of stars with discs reduces greatly with increasing time \citep{2009AIPC.1158....3M}. In our nominal model, the lifetime of the disc is fixed to $3$ Myr, where we follow the disc evolution of \citet{1998ApJ...495..385H}. At $3$ Myr the disc has reached a stellar accretion rate of $\dot{M} = 2 \times 10^{-9} M_\odot/$yr. We then assume that the disc gets cleared by photoevaporation, so we stop our simulations at that time. However, the effects of photoevaporation are not clear and not constrained perfectly \citep{2013arXiv1311.1819A}, so that we now make the assumption that the disc will live longer, up to $5$ Myr, where the disc will reach an accretion rate of $\dot{M} = 1 \times 10^{-9} M_\odot/$yr \citep{1998ApJ...495..385H}. The additional lifetime of $2$ Myr therefore only models the decay of the accretion rate from $\dot{M} = 2 \times 10^{-9} M_\odot/$yr to $\dot{M} = 1 \times 10^{-9} M_\odot/$yr.

\begin{figure}
 \centering
 \includegraphics[scale=0.7]{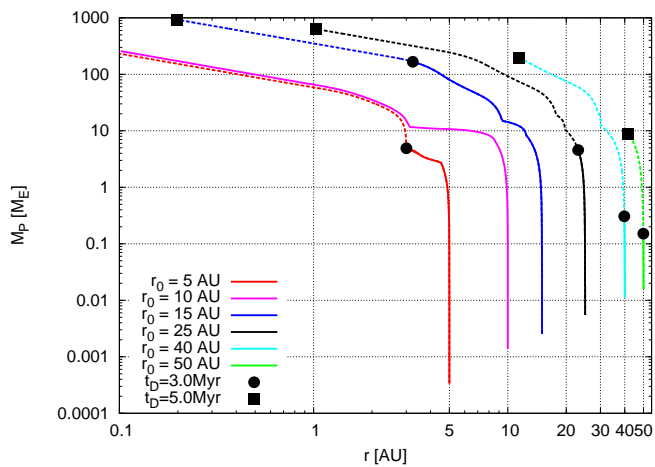}
 \caption{Evolution of planets that start at $t_0=2$ Myr, but in discs that live up to $t_D=3$ Myr or $t_D=5$ Myr. The black circular symbols mark $t_D=3$ Myr, which corresponds to the final positions in Fig.~\ref{fig:Envhockey}. The black squares mark the final position at $t_D=5$ Myr. The evolution between $3$ Myr and $5$ Myr is indicated by the dashed lines. Clearly planets in the disc with longer lifetimes continue their evolution, so that the final dissipation time of the disc is very important to set the final mass and orbital position of planets.
   \label{fig:Hockey5Myr}
   }
\end{figure}

Starting from the same initial configuration as for Fig.~\ref{fig:Envhockey}, namely starting planets at different orbital distances when the disc has already evolved to $t_{\rm D} = 2$ Myr, results in Fig.~\ref{fig:Hockey5Myr}, where we present the growth tracks of planets in discs that live $3$~ Myr and $5$ Myr. The planets are inserted at $t_D=2$ Myr, so they undergo $1$~Myr or $3$ Myr of evolution, respectively. The evolution of the planets in the disc that lives longer is identical to the disc that lives for a shorter time, because both discs have the same properties.

In the disc that lives longer, planets have more time for their evolution. This means that the planets, which have not contracted their gaseous envelope completely, can now do so and thus start rapid gas accretion. This is clearly seen for the planet in the inner system (Fig.~\ref{fig:Hockey5Myr}), which classified as an ice giant in the $3$ Myr disc, while it is a hot gas giant in the disc that lives for $5$ Myr. Similar fates are shared by the planets that form in the outer system. Those who already started rapid gas accretion accrete so much gas that the planet becomes massive enough that type-II migration shifts from disc-dominated to planet-dominated (eq.~\ref{eq:typeII}), reducing its migration speed. The ice giants in the outer disc now have more time to accrete pebbles and reach their isolation mass, so that they can contract their envelope and undergo runaway gas accretion, transforming them into gas giants. Even the planetary seed that would only grow to become an ice planet in a disc that lives $3$ Myr (Fig.~\ref{fig:Envhockey}) evolves into a gas giant. This also indicates that the lifetime of the disc does not only play any role in terminating the growth of gas giants, but for planets in general.

\begin{figure}
 \centering
 \includegraphics[scale=0.69]{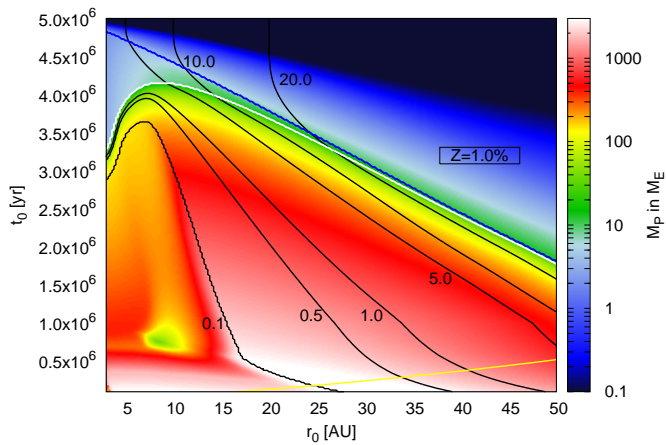}
 \caption{Final masses of planets as a function of initial radius $r_{\rm 0}$ and initial time $t_{\rm 0}$ in the disc, where $Z=1.0\%$, and a total disc lifetime of $5$ Myr. The lines inside the plots correspond to the same meaning as the lines in Fig.~\ref{fig:Z008kappa}. Clearly the formation of small planets is delayed to times after $t_0=3.5$ Myr.
   \label{fig:Z008K0005Myr5}
   }
\end{figure}

In Fig.~\ref{fig:Z008K0005Myr5} we present the $r_0$-$t_0$ map for a disc that undergoes $5$ Myr of evolution. The general shape of the different regions where different planetary types emerge (e.g. gas giants, ice giants) is very similar to Fig.~\ref{fig:Z008kappa} where the disc lived for $3$ Myr in total. The difference is now that the formation of all planetary types can start at $3.5$ Myr, compared to $\approx 2$ Myr in the disc that lives $3$ Myr. Otherwise, the general results do not change. A longer disc lifetime transforms smaller planets to larger ons, because they have more time to grow, and reduces the final semi-major axis of those planets, because they have more time to migrate.

\section{Constraints on the opacity during envelope contraction}
\label{ap:kappa}

After the core reaches pebble isolation mass, it hosts a gaseous envelope that first undergoes contraction on a long time scale, before runaway gas accretion can start. During this contraction phase, the envelope grows slowly until $M_{\rm c} = M_{\rm env}$, which is when rapid gas accretion can start. The accretion rate during the contraction phase is given by eq.~\ref{eq:Mdotenv}. The opacity $\kappa_{\rm env}$ inside the envelope is crucial in determining the accretion rate during envelope contraction. Unfortunately, the opacity inside the planets envelope is poorly constrained \citep{2000ApJ...537.1013I}. Not only is the size of the grains important for determining the opacity, but also the temperature in the planetary atmosphere. As grain growth can be efficient in planetary atmospheres \citep{2014A&A...572A.118M} and the temperature of a planet accreting gas is higher than the surrounding disc, the opacity of the planetary atmosphere is reduced compared to the opacity in the disc. Additionally, the opacity depends on the underlying chemical composition of the grains, which depend on the composition of the disc and the temperature in the disc. 

\begin{figure}
 \centering
 \includegraphics[scale=0.69]{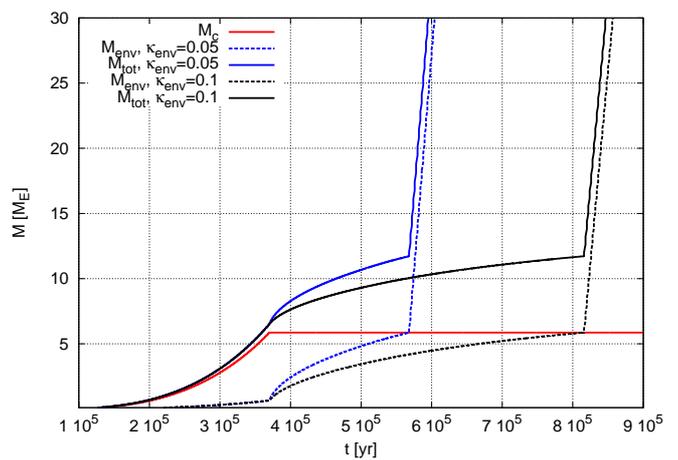}
 \caption{Mass of the planet as a function of time. The mass of the core (red line) increases via pebble accretion until isolation mass is reached at $350$ kyr. Then the envelope starts to contract (dashed lines), which in increases the total mass (solid lines). The contraction time of the envelope depends on the opacity of the envelope $\kappa_{\rm env}$, where lower opacity results in a shorter contraction time, so that when $M_{\rm c} = M_{\rm env}$, rapid gas accretion can start. The planet is placed in a disc at $t_0=2$ Myr at $r_0=10$~AU.
   \label{fig:gasaccrete}
   }
\end{figure}

The nominal opacity in eq.~\ref{eq:Mdotenv} is set to $\kappa_{\rm env}=0.05$ ${\rm cm}^2/{\rm g}$. Further reductions in the opacity of the envelope result in faster contraction times of the envelope, so that $M_{\rm c}<M_{\rm env}$ is reached at shorter times, allowing runaway gas accretion during the disc's lifetime. In fact, a close-to-zero opacity allows rapid gas accretion onto cores of just $1 {\rm M}_{\rm E}$ \citep{2010ApJ...714.1343H}. When setting $\kappa_{\rm env}=0.05$ ${\rm cm}^2/{\rm g}$, the contraction time of the envelope is roughly of the as the building time of the core via the pebble accretion shown in Fig.~\ref{fig:gasaccrete}. This value is very close to the estimate of opacity in the envelope of \citet{2008Icar..194..368M}.

In Fig.~\ref{fig:gasaccrete} we display the growth of a planet as a function of time. After the core has reached pebble isolation mass, the envelope starts to contract, where the opacity in the envelope $\kappa_{\rm env}$ sets the contraction time. It clearly shows that higher opacity in the envelope prolongs the contraction time of the envelope.

A prolongation of the contraction time has important consequences for the formation of planets, because the planet will then spend a longer time in type-I migration before it is massive enough to open a gap in the disc. This means that the distance the planet migrates from its initial position $r_{\rm 0}$ is greater for a higher values $\kappa_{\rm env}$, unless the planet is caught in a region of outward migration. However, this can only happen in the early evolution stages of the disc. In the late stages of the disc evolution, a longer contraction time will lead to an enlargement of parameter space that allows for the formation of ice giants at a few AU, because these planets have very small cores and therefore do not contract an envelope.

\section{Planet formation in the MMSN}
\label{ap:MMSN}

\begin{figure}
 \centering
 \includegraphics[scale=0.69]{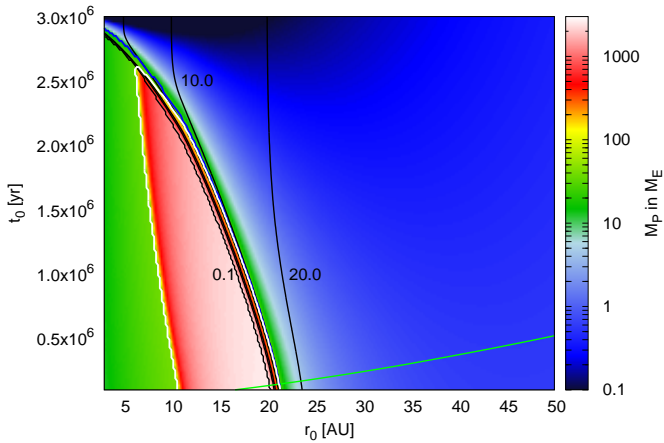}
 \caption{Final masses of planets as a function of initial radius $r_{\rm 0}$ and initial time $t_{\rm 0}$ in the MMSN with a lifetime of $3$ Myr. The lines inside the plots correspond to the same meaning as the lines in Fig.~\ref{fig:Z008kappa}. We again use a metallicity of pebbles of $Z=1.0\%$. In contrast to the disc model of \citet{2014arXiv1411.3255B}, the MMSN disc allows $0.1$~AU $<r_{\rm f}<5.0$~AU only for a very small amount of parameter space. Either the planets fall close to the central star ($r_{\rm f}<0.1$~AU) or they stay outside of $5$~AU. On the other hand, the parameter region that allows for the formation of ice giants is greatly increased compared to Fig.~\ref{fig:Z008kappa}.
   \label{fig:MMSN}
   }
\end{figure}

To emphasize the importance of a realistic disc structure for the formation of planets, we tested our planet formation model also with a MMSN disc \citep{1977Ap&SS..51..153W, 1981PThPS..70...35H}. A simple power law disc was also used in \citet{2014arXiv1408.6094L}, but with slightly different parameters compared to the MMSN. Here we directly probe the MMSN, where the discs surface density is given by
\begin{equation}
 \Sigma_{\rm g} (r) = \beta_\Sigma \left(\frac{r}{\rm AU}\right)^{-3/2} \ .
\end{equation}
The aspect ratio of the disc follows
\begin{equation}
 \frac{H}{r} = 0.033 \left(\frac{r}{\rm AU}\right)^{1/4} \ ,
\end{equation}
which indicates that there are no planet traps due to the entropy driven corotation torque, because those traps exist only when $H/r$ decreases with radius \citep{2014A&A...564A.135B}. The time evolution of the surface density is given by
\begin{equation}
 \beta_\Sigma = \Sigma_0 \exp \left(\frac{-t}{\tau_{\rm disc}}\right) \ ,
\end{equation}
where $\tau_{\rm disc} = 3$ Myr and $\Sigma_0 = 1700 {\rm g}/{\rm cm^3}$. From this disc structure we model the formation of planets, and the resulting $r_0$-$t_0$ diagram is shown in Fig.~\ref{fig:MMSN}.

The resulting planets look dramatically different from those in Fig.~\ref{fig:Z008kappa}. In the inner regions of the disc ($r_0<10$~AU), the planets do not reach rapid gas accretion, because the migration in the inner disc is so rapid (because of the high surface density) that the planets arrive at $r_{\rm f}<0.1$~AU before $M_{\rm c}<M_{\rm env}$. Planets that form farther outside can reach this stage ($M_{\rm c}<M_{\rm env}$), but their migration is still so rapid that they end up at $r_{\rm f}<0.1$~AU. Nearly a third of the $r_0$-$t_0$ parameter space results in planets with $r_{\rm f}<0.1$~AU. Only a very small band allows for the formation of planets with $0.1$~AU $<r_{\rm f}<5.0$~AU, where most of these planets have $M_{\rm c}>M_{\rm env}$, indicating that the formation of ice giants outside of $1$~AU and inside of $5$~AU is very hard in this disc model. 

However, outside of $5$~AU, a very large band allows for the formation of ice giants (green region in Fig.~\ref{fig:MMSN}). These ice giants have a much larger number of solids ($\sim 20-30 {\rm M}_{\rm E}$) compared to Fig.~\ref{fig:Z008kappa}. This is caused by the higher aspect ratio in the MMSN disc, compared to our nominal disc model \citep{2014arXiv1411.3255B}, which increases the pebble isolation mass (eq.~\ref{eq:Misolation}). Additionally, in the MMSN disc model, only the surface density evolves in time, while the aspect ratio does not evolve, allowing a high pebble isolation mass in all disc evolution stages. In combination with the efficient pebble accretion, planets in this disc model will have a higher core mass than in the nominal disc model used in Fig.~\ref{fig:Z008kappa}.

\section{The heating torque}
\label{ap:heating}

A recent study by \citet{2015Natur.520...63B} shows that a small planet ($M_{\rm P} < 5 {\rm M}_{\rm E}$) that accretes very quickly (mass doubling times shorter than $60$ kyr) alters the structure of the surrounding gas disc in such a way that it can migrate outwards in the disc. The accreting material onto the planet causes asymmetries in the temperature of the disc close to the planet that produce a force that counteracts inward migration. This effect concerns planets that are so small that they had not yet reached the region of outward migration (see Fig.~\ref{fig:Migcont}). Even if the mass doubling time is large, e.g. $300$ kyr, the heating torque has an effect, because it reduces the inward speed of the planet compared to the nominal type-I migration rate.

This so-called {\it \emph{heating torque}} is a strong function of the accretion rate onto the planet (only mass doubling times shorter than $60$ kyr lead to outward migration) and of the opacity of the disc, which determines the cooling in the surrounding disc. Higher opacity suppresses cooling, which leads to a stronger effect of the heating torque. In their standard set-up, \citet{2015Natur.520...63B} use a constant opacity of $1$ ${\rm cm}^2/{\rm g}$, and the planet has a mass of $3$ ${\rm M}_{\rm E}$. The effect of the heating torque also depends on the mass of the planet. Unfortunately, this new effect is not quantified in a function that contains all relevant parameters (planetary mass, opacity of the disc, and mass doubling time). We therefore make a simple test, where we assume that the heating torque is just a function of the accretion rate and do not investigate effects of opacity and planetary mass in a detailed way. For that we use the data in the methods section of \citet{2015Natur.520...63B}.

We assume that the effect of the heating torque is equally strong until the planet reaches $5$~${\rm M}_{\rm E}$, after which the heating torque is not relevant any more. Additionally, we assume just for the calculation of the heating torque that the opacity is equal to $1$ ${\rm cm}^2/{\rm g}$, which is actually not the case in our disc. In fact, only in the inner parts of the disc, so close to the ice line, is the opacity higher than $1$ ${\rm cm}^2/{\rm g}$ in our disc model. Depending on the initial semi-major axis $r_0$ and the initial time $t_0$, the mass doubling time in our simulations can be a few kyr (small planet with $\sim 0.5$ ${\rm M}_{\rm E}$, small $r_0$, and small $t_0$) up to a few $100$ kyr (large planet, larger $r_0$, and larger $t_0$).

\begin{figure}
 \centering
 \includegraphics[scale=0.69]{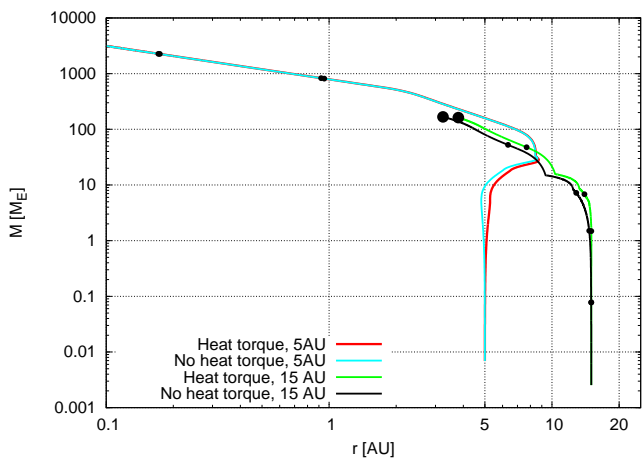}
 \caption{Two examples of the evolution of planets with and without heating torque. Both planets starting at $5$~AU start at an early initial time $t_0=100$ kyr, while the planets starting at $15$~AU start at a late initial time of $t_0=2$ Myr. The small black dots indicate a time difference of $200$, $400$, $600,$ and $800$ kyr from the starting time. The big black dots indicate an evolution time of $1$ Myr. The planets starting early and in the inner disc only evolve for $\sim 500$ kyr total. With the fast growth rates of pebble accretion, it seems that the heating torque only has minimal effects on the final configuration of evolving planets.
   \label{fig:heathockey}
   }
\end{figure}

In Fig.~\ref{fig:heathockey} we display the evolution tracks of planets with and without the heating torque. In the example where the planets start at $5$~AU, they are also inserted into the disc at an early time ($t_0=100$ kyr). In the early disc, the region of outward migration is still very large (Fig.~\ref{fig:Migcont}), so that planets that are massive enough (a few Earth masses) to reach this region and can grow there without migrating inwards. In this example, because $r_0$ and $t_0$ are small, the heating torque easily generates outward migration, and the planet ends up in the region of outward migration. However, the planet also reaches the region of outward migration when the heating torque is not taken into account, because the growth by pebble accretion is so fast that the planet does not migrate much during this time. Both planets then end up on the same evolution track, because they reach the region of outward migration where their previous migration history no longer matter.

The planets starting at $20$~AU in Fig.~\ref{fig:heathockey} also start later in the disc at $t_0=2$ Myr. There the planets do not reach a region of outward migration. Additionally, the mass doubling time becomes longer than $60$ kyr for most of the growth stages of the planets, so that the heating torque does not prevent inward migration, but just slows it down. However, even this slowing down has an effect on the final orbital mass and position of the planet. The planet that grows when we take the heating torque into account has a somewhat larger final semi major axis, but reaches a similar final planetary mass. The reason for this effect not seeming too important here is that small mass bodies migrate quite slowly, and they also grow quite quickly thanks to pebble accretion, so that they reach $5$ ${\rm M}_{\rm E}$ rather quickly, so that the planet experiences the effects of the heating torque for only a small amount of time.

The general results of our simulations, for example Fig.~\ref{fig:Z008kappa}, still holds, even when taking the heating torque into account. The only thing that changes is that the final orbital positions $r_f$ of the planets are slightly larger, meaning that a smaller $r_0$ is needed to achieve a certain final orbital position $r_f$. In the simple test here, the initial semi-major axis can be up to $2$~AU smaller in order to achieve the same $r_f$ for a given $t_0$ compared to simulations where the heating torque is not taken into account, making the effect in a global picture not that important here.

This effect seems to not be too important in the pebble accretion scenario because of the fast growth rates. The fast growth rates for reaching $M_{\rm P}>5{\rm M}_{\rm E}$ mean that there is only a short time for the planet to actually spend migrating. The resulting difference between simulations with and without heating torque where the accretion rates are high is therefore minimal. However, the heating torque could have substantial effects when the accretion rate is low, and the heating torque reduces the torque causing inward migration. In \citet{2014arXiv1407.6011C} the growth of planetary cores is modelled in N-body simulations via the collision of planetesimals. There the production of the cores of giant planets at large orbits is hindered by the fact that planets migrate inwards too fast below the region of outward migration (because the planets are not massive enough and do not grow fast enough), which results in many super-Earth types of planets instead of giant planets. Including the heating torque this picture could change, because the inward migration speed is reduced for these small objects, which could allow them to grow more and reach the region of outward migration. Additionally, if the metallicity caused by $\mu$m sized dust grains is greater than the $0.5\%$ as in our work, the heating torque could become more prominent, because it strongly increases with increasing opacity, which directly scales with the metallicity.

\section{Type-III migration and dynamical corotation torque}
\label{ap:typeIII}

When a planet is big enough to carve a gap around its orbit, the horseshoe region becomes depleted. However, when the mass pushed away by the planet is higher than the mass of the planet itself (the co-orbital mass deficit), the planet can undergo a rapid change of semi-major axis, which is described as type-III migration \citep{2003ApJ...588..494M}. Migrating planets also experience dynamical torques, which are proportional to the migration rate and depend on the background vortensity gradient \citep{2014MNRAS.444.2031P}. Here we want to discuss how these effects influence our models. The description of how these mechanisms work can be found in section~\ref{subsec:migration}.

\begin{figure}
 \centering
 \includegraphics[scale=0.69]{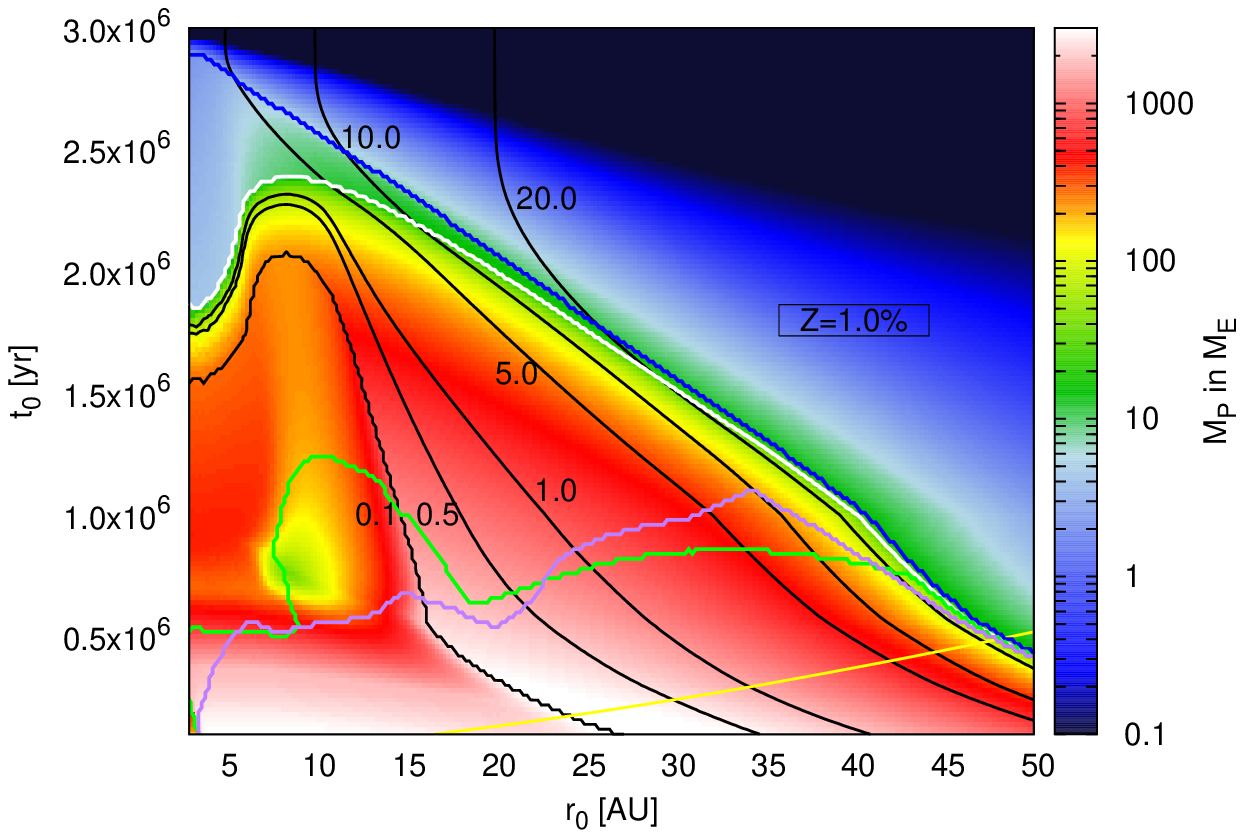}
 \includegraphics[scale=0.69]{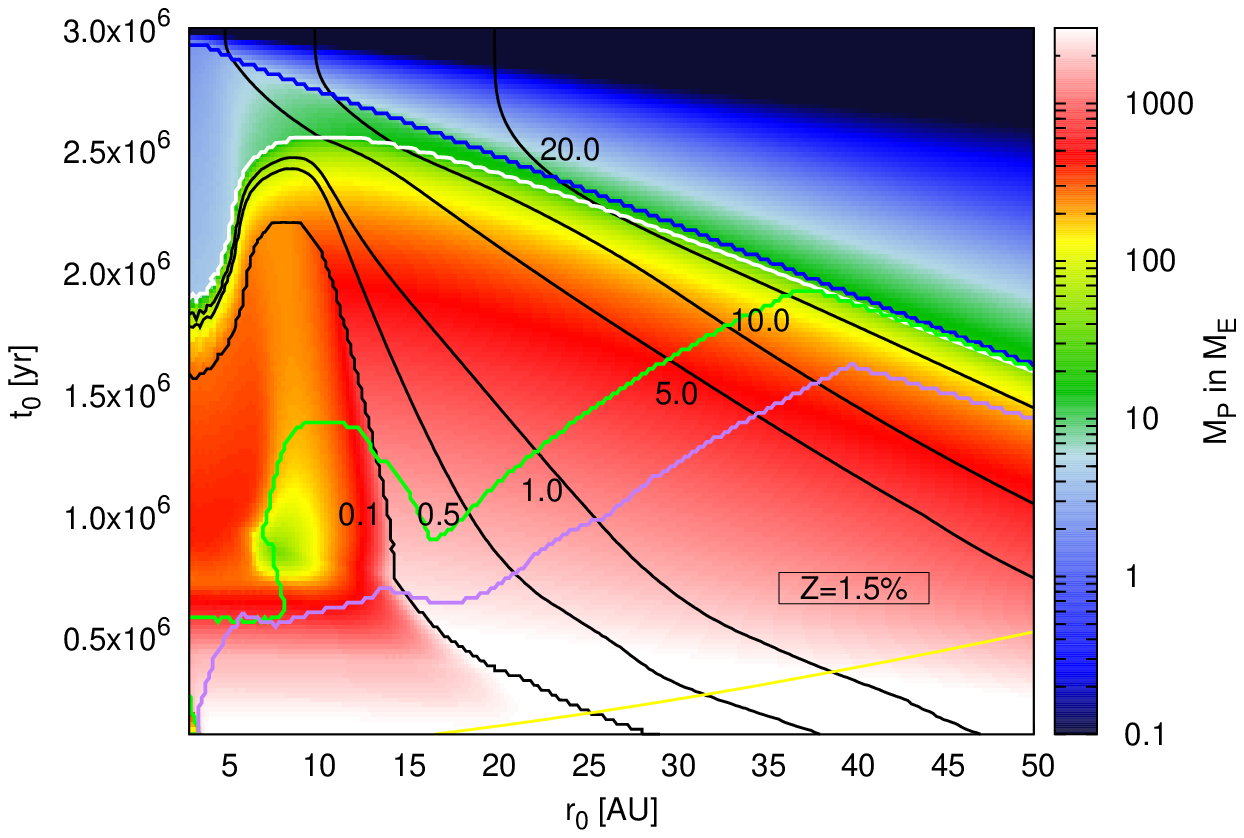}
 \caption{Final masses of planets as a function of initial radius $r_{\rm 0}$ and initial time $t_{\rm 0}$ in the disc with $Z=1.0\%$ (top) and $Z=1.5\%$ (bottom). The lines inside the plots correspond to the same meaning as the lines in Fig.~\ref{fig:Z008kappa}. Planets below the purple line are subject to dynamical corotation torques \citep{2014MNRAS.444.2031P}, while planets below the green line are subject to type-III migration \citep{2003ApJ...588..494M}, indicating that their final orbital position does not correspond exactly to the position indicated by the black lines. 
   \label{fig:Z008paartype3}
   }
\end{figure}

In Fig.~\ref{fig:Z008paartype3} we have indicated the planets that would either undergo type-III migration or are subject to dynamical corotation torques (everything below the green and purple lines in Fig.~\ref{fig:Z008paartype3}) resulting in the fact that the final orbital positions in these regions are not exactly marked by the black lines. The final orbital positions are most likely closer to the host star, because the runaway type-III migration follows the direction of motion, which is inwards in regions outside of $r>12$~AU (Fig.~\ref{fig:Migcont}). This process only occurs in the early stages of the disc evolution, where the disc is still massive. In the later stages of the disc evolution, the disc is less massive, and these effects are not significant any more, making the predictions of our model reliable in this part of parameter space.

Simulations with a larger number of pebbles can form gas giants more easily and in a larger parameter space in $r_0$-$t_0$ (Fig.~\ref{fig:Zenvelope}). A larger number of pebbles results in a faster growth of the planetary core, which then can contract its envelope faster as well, leading to a total faster growth of the planet. This means the planet will reach the mass where it starts to open a gap partially in the disc earlier as well. However, in an earlier stage of evolution, the disc is also more massiv, indicating that the co-orbital mass deficit is larger as well, making the planet more prone to type-III migration. The same effect applies for the dynamical corotation torque. In the situation of a larger number of pebbles in the disc, a greater number of planets in the $r_0$-$t_0$ parameter space are affected by type-III migration and dynamical corotation torques (Fig.~\ref{fig:Z008paartype3}).

\bibliographystyle{aa}
\bibliography{Stellar}
\end{document}